\documentclass[11pt,letterpaper]{article}
\usepackage[T1]{fontenc}
\usepackage{amsmath,amsthm}
\usepackage{newtxtext,newtxmath}

\usepackage[margin=1in,headheight=14pt]{geometry}
\usepackage{setspace,fancyhdr}
\makeatletter
\renewcommand{\section}{\@startsection{section}{1}{0pt}%
  {-2.1ex plus -.5ex minus -.3ex}{.8ex}{\large\sffamily\bfseries}}
\renewcommand{\subsection}{\@startsection{subsection}{2}{0pt}%
  {-1.6ex plus -.4ex minus -.2ex}{.5ex}{\normalsize\sffamily\bfseries}}
\renewcommand{\@seccntformat}[1]{\csname the#1\endcsname.\enspace}
\makeatother
\usepackage{booktabs,array,tabularx}
\usepackage{algorithm,algpseudocode}
\usepackage[numbers,sort&compress]{natbib}
\usepackage[hidelinks]{hyperref}
\hypersetup{pdftitle={Achieving 1-1/e in Online Matching with Unit-Capacity Reusable Resources},
  pdfauthor={Yilin Cheng},pdfsubject={Preprint},
  pdfkeywords={online matching, reusable resources, competitive analysis, randomized algorithms}}
\newtheorem{theorem}{Theorem}
\newtheorem{lemma}{Lemma}
\newtheorem{proposition}{Proposition}
\newtheorem{corollary}{Corollary}
\newtheorem{conjecture}{Conjecture}
\makeatletter
\renewcommand{\@maketitle}{%
  \begin{singlespace}
    \raggedright
    {\LARGE\sffamily\bfseries \@title\par}
    \vspace{8pt}
    {\large \@author\par}
  \end{singlespace}}
\makeatother
\title{Achieving $1-1/e$ in Online Matching with Unit-Capacity Reusable Resources}
\author{Yilin Cheng\textsuperscript{a,b}\\[2pt]
  \small\textsuperscript{a}Fudan University, Shanghai, China\qquad
  \small\textsuperscript{b}Zhongguancun Academy, Beijing, China}
\date{}

\begin{document}
\maketitle
\begin{singlespace}
\small
\noindent\textbf{Abstract.} We study online bipartite matching with one unit of each resource. Assigning a resource earns its known weight and makes it unavailable for a common deterministic duration. Delong et al.\ achieved a competitive ratio of $58.9\%$ and asked whether $1-1/e$ is attainable. We give a randomized policy whose expected reward on every fixed input is at least $1-1/e$ (approximately $63.21\%$) times the optimum of the fractional window relaxation. When a resource loses a bid, the policy suspends the ranks that would have won and restores them one service duration later. This timing lets the analysis charge a request to an earlier match of a resource that is now busy. The ratio is optimal uniformly over durations. With at most one request per integer slot and duration $d$, the same policy guarantees $1-1/e+2^{-21}/d^2$. At $d=2$, a separate policy attains the optimal worst-case ratio of $3/4$.
\par
\smallskip
\noindent\textbf{Keywords.} Online matching, reusable resources, competitive analysis, randomized algorithms.
\par\smallskip\hrule
\end{singlespace}
\section{Introduction}
An online scheduler has one unit of each resource. Requests arrive one at a time, each revealing the resources that can serve it. Assigning a free resource earns its known weight and occupies it for a common deterministic duration $d$. The request sequence is fixed independently of the scheduler's randomness, and each decision must be made before the next request is revealed. Delong et al.\ \cite{delong} obtained a competitive ratio of $58.9\%$ for this problem. In their words, ``The main open problem left unanswered in our paper is whether one can obtain the competitive ratio of $1-1/e$ for OBMRR'' \cite[Section~5, p.~1849]{delong}.

We answer this question with a randomized policy called \emph{delayed restoration}. On every fixed input, its expected reward is at least $1-1/e$ times the value of the full window LP (Theorem~\ref{thm:main}). The LP allows fractional assignments and bounds the total use of each resource in every window of $d$ slots. Its value can be $7/2$ on an input whose best offline schedule earns only three \cite[Remark~1]{delong}. The policy needs only the free eligible resources, their known weights, and the common clock.

\paragraph{Contributions.}
\emph{The uniform ratio.}
Delayed restoration attains $1-1/e$ for every positive integer duration, even when menus change arbitrarily and the horizon is unknown. When the duration covers the whole input, resources cannot return and the classical upper bound for ordinary matching applies. With arbitrary real arrival times, that instance can fit before the first return at any prescribed positive duration (Corollary~\ref{cor:real}).

\emph{The window coupling.}
If a resource is busy at a target request, the proof must account for an assignment made earlier. Theorem~\ref{thm:window-coupling} compares two runs on the same random paths, one with the target resource and one without it. Whenever the resource's potential bid exceeds the terminal price in the comparison run, the actual run used the resource within the window. That assignment supplies the resource credit.

\emph{Fixed integer durations.}
With at most one request per slot, the same policy gains at least $2^{-21}/d^2$ above $1-1/e$ against the full window LP (Theorem~\ref{thm:untrimmed}). The optimal ratio is nonincreasing in $d$ (Proposition~\ref{prop:duration-order}). At $d=2$, a policy that sees the complete current menu attains the optimal ratio for each weight vector; the worst of these ratios is $3/4$ (Theorem~\ref{thm:fixed-two}).

\paragraph{What information survives a losing bid?}
Perturbing the known weights by independent random ranks is effective for ordinary online matching. Reuse creates an additional dependence: a resource that loses remains free, and the winning price reveals information about its private rank. Drawing new independent ranks without accounting for this information loses the structure used by a critical-rank comparison.

Delayed restoration keeps a region of active ranks for each resource. After that resource loses to price $b$, the policy suspends the active ranks that would have beaten $b$. It restores precisely this region one service duration later. When a physical use finishes, the resource instead restarts from the full rank interval. The rank drawn on an active region is the largest point of a Poisson process with intensity $dz/z$ there. Restricting this process below a revealed cutoff gives the posterior distribution after a loss. Adding an independent process on a returning region gives the restored distribution. These two identities provide a persistent representation in which an unused resource can retain its private rank between requests.

\paragraph{A window coupling.}
For a target request $t$, set the boundary at $r=\max\{0,t-d\}$. A suspension made before or at this boundary expires by $t$, whereas one made strictly inside the window survives until after $t$. For $t>d$, the boundary and the first and last suspension times in the window have the following deadlines:
\[
\begin{array}{r@{\qquad}c@{\quad}c}
 &\text{boundary}&\text{comparison window }W_t\\
 \text{slot}&r=t-d&r+1\ \longrightarrow\ \cdots\ \longrightarrow\ t\\
 \text{restoration slot}&t&t+1\ \longrightarrow\ \cdots\ \longrightarrow\ t+d
\end{array}
\]
Conditioning on the winners and prices through $r$ therefore allows an independent potential rank path for each resource to be fixed at the boundary. Its terminal rank at $t$ is uniform. A rank region suspended inside the window contains no point that has already been revealed. Consequently the actual rank of an unused resource agrees with the maximum of the revealed points along its potential path (Lemma~\ref{lem:window-invariant}).

Run the deterministic decision map on these paths, then run the same map with a target resource $i$ deleted. The terminal rank of $i$ is independent of the deleted run. If its terminal bid exceeds the deleted run's terminal price, the actual run must use $i$ somewhere in the window. Its rank at that earlier use is no greater than its terminal rank. Splitting a realized reward between the current request and the selected resource then gives a lower bound for the target edge, even when $i$ is busy at $t$. Multiplying these inequalities by a feasible window allocation and summing charges each request payment and each resource credit at most once. Section~\ref{sec:continuous} proves these statements on a common probability space for each target edge.

\paragraph{The strict gain at fixed duration.}
The baseline charge does not use all available reward. A resource used before the target can leave credit beyond the amount required by the baseline, while a sufficiently large terminal price can itself improve the request payment. The strict-gain proof separates the price comparisons where neither contribution is large. Deleting a second resource controls their probability. An arbitrary menu may contain many competitors, but at most $2d$ resources can each carry half a unit of comparator mass in the forward window. When the winner falls outside this set, its remaining credit pays for the exceptional comparison. This gives a bound depending on the window width and hence the gain $2^{-21}/d^2$. The proof uses the comparator to account for reward, while the algorithm uses its original menus and private randomness.

The factor $d^{-2}$ comes from the price-overshoot estimate in this proof. The classical permutation instance gives an $O(1/d)$ upper bound on the excess of every policy over $1-1/e$. Determining the best order at fixed integer duration remains open. Duration two can be solved exactly by a policy that uses the complete revealed menu to track the offline prefix value. For weights $(1,2)$, this policy guarantees $7/9$, whereas delayed restoration obtains about $0.73644$ on a two-request input. Thus the policy that resolves the uniform question need not optimize every fixed-duration problem.

\paragraph{Prior guarantees and benchmarks.}
Table~\ref{tab:prior-guarantees} places the unit-inventory guarantee alongside results under other assumptions on inventory and usage times. All rows allow a request sequence fixed independently of the algorithm's random choices. The stochastic-duration benchmarks know future requests and choose a use before seeing its realized duration. Goyal et al.\ \cite{giu} call their benchmark clairvoyant with respect to the arrival sequence and assume large starting capacities. Bei et al.\ \cite{bei} study unit inventory. Our comparison is with the full fractional window LP at unit inventory.
\begin{table}[t]
\caption{Competitive guarantees for reusable resources. In the deterministic rows, the common duration is known.}\label{tab:prior-guarantees}
\small
\begin{tabularx}{\linewidth}{@{}>{\raggedright\arraybackslash}p{.20\linewidth}>{\raggedright\arraybackslash}p{.23\linewidth}>{\raggedright\arraybackslash}p{.24\linewidth}>{\raggedright\arraybackslash}X@{}}
\toprule
Method & Inventory and durations & Required information & Guarantee and benchmark\\
\midrule
Weight-greedy \cite[Section~1]{delong} & Unit, common deterministic & Resource weights and current free offers & $1/2$ against the window LP\\
Periodic Reranking \cite[Theorem~1]{delong} & Unit, common deterministic & Resource weights, free offers and common clock & $0.589$ against the window LP\\
Goyal et al.\ \cite[Theorem~2, full version]{giu} & Large capacities, independent resource-specific durations & Weights, capacities and usage distributions & Asymptotically $1-1/e$ against clairvoyant OPT\\
Fixed-order Ranking \cite[Theorem~1.1]{bei} & Unit, resource-specific i.i.d. durations, unweighted & Free offers, no distribution oracle & $(5-2\sqrt3)/3$ against nonanticipatory OPT\\
Delayed restoration (Theorem~\ref{thm:main}) & Unit, common deterministic & Resource weights, free offers and common clock & $1-1/e$ against the window LP\\
\bottomrule
\end{tabularx}
\end{table}

The greedy bound in Table~\ref{tab:prior-guarantees} also holds for the fractional benchmark. For each eligible pair $(i,t)$, either $i$ is free and the greedy reward at $t$ is at least $w_i$, or $i$ is busy and earned $w_i$ at an earlier slot in $W_t$. Multiply by $x_{it}$ and sum. By the two constraints in \eqref{eq:window-lp}, each realized reward is counted with coefficient at most one as a request reward and at most one as a resource reward. Hence $\mathrm{LP}\le2\,\mathrm{Greedy}$.

\paragraph{Ranking and correlated selection.}
The vertex-weighted perturbation of Aggarwal et al.\ \cite{agkm} and the critical-rank comparison of Devanur et al.\ \cite{dj} give the starting point for our analysis. Liang et al.\ \cite{liang} characterize optimal perturbations in ordinary matching. If $E$ is a rate-one exponential random variable, then $e^{-E}$ is uniform on $(0,1)$. The Poisson construction in Section~\ref{sec:continuous} extends this representation to restrictions on changing active regions. The new step is to restore information discarded after a losing comparison at the time required by a reusable-resource window.

Fahrbach et al.\ \cite{fhtz} introduced online correlated selection to obtain an improvement over $1/2$ in edge-weighted online matching with free disposal. Delong et al.\ \cite{delong} develop an online correlated rental construction for reusable resources. Their journal article merges the EC~2022 work on this construction \cite{delongEC} and on Periodic Reranking \cite{udwaniEC}. Our policy instead redraws private ranks conditionally on their active regions and uses the reward of an earlier match to account for a later request. In fully online matching, Huang et al.\ \cite{fully} share gains beyond the endpoints of a matched edge. Reusability requires accounting for repeated uses of the same resource.

A preliminary note \cite{partial} studies blocks that span at most $d$ slots, with the first request of each new block arriving at least $d$ slots after the last request of its predecessor. Within each block a resource can be used at most once, and every resource is free at the start of the next block. The offline optimum therefore adds across blocks. Running vertex-weighted Perturbed Greedy \cite{agkm} with fresh ranks in each block gives $1-1/e$ by the ordinary matching guarantee. For general menus, assignments in one block can still occupy resources in the next. The coupling in Section~\ref{sec:continuous} accounts for precisely those earlier assignments.

\paragraph{Fixed ranks and ranks redrawn on return.}
The Ranking algorithm of Bei et al.\ \cite{bei} draws one priority order before the first request and keeps it throughout the entire input. It earns $(5-2\sqrt3)/3$ of their unweighted stochastic-duration benchmark without knowing the duration distributions. Their weighted reduction \cite[Corollary~1.3]{bei} gives a guarantee of $0.500034$ with independent sampling access to those distributions. Reranking on Return draws a new rank after each completed use and retains a losing free resource's rank. Thus one policy retains an order across uses, and the other redraws it on return. Delayed restoration also changes the active rank region after a loss.

\paragraph{Degree restrictions and stochastic allocation.}
Naor and Wajc \cite{nw} bound the online degree above and every offline degree below. Cohen and Wajc \cite{cw} obtain ratios approaching one on regular graphs. Our degree bound restricts only the arriving menus. For unweighted ordinary Ranking, Xu \cite[Section~7.2]{xu} proves the stronger gain $1/[3e\Delta(\Delta+2)]$ at online degree $\Delta\ge2$. Bhangale et al.\ \cite{bch} obtain distinct optimal ratios for fractional and randomized integral matching on unweighted inputs of online degree at most two. The second deletion in our proof yields a strict gain with arbitrary resource weights and returns.

Stochastic reusable-resource allocation has also been studied through dynamic assortment \cite{rst}, myopic assortment under adversarial preferences \cite{gong}, and resource bundles \cite{bm}. Bayesian large-inventory policies appear in Feng et al.\ \cite{fns}, while Dickerson et al.\ \cite{dssx} assume known arrival and occupation-time distributions. Baek and Wang \cite{bw} give the greedy guarantee $1/(2-p)$, where $p$ is the minimum next-slot return probability. Here $p=0$ for deterministic $d>1$. Liu and Hajiesmaili \cite{liu} study reusable configurations under deterioration. Ekbatani et al.\ \cite{jobs} allow job--server-specific rewards and durations as minimum capacity grows. Our main theorem uses one unit of each resource and a common deterministic duration.

\section{Model and guarantee}\label{sec:model}
Let $I$ be a finite resource set with known nonnegative weights $w_i$, and let $d\ge1$ be a known integer duration. At slot $t$, a request reveals an eligibility menu $N_t\subseteq I$. A resource is free if all of its earlier uses have finished. The policy may assign one free member of $N_t$ or reject the request. An assignment to $i$ at slot $s$ earns $w_i$. The resource returns before the decision at slot $s+d$. Rejection earns zero. All menus and the finite termination time are fixed independently of the private random draws. The policy receives no advance knowledge of future menus or of the termination time.

We use \emph{resource} for a member of $I$, \emph{request} for one arrival, \emph{menu} for the input set $N_t$, and \emph{free menu} for its currently available members. Algorithms that receive only the free menu and algorithms that receive the complete menu are distinguished in Table~\ref{tab:information}. Resource names are used only to resolve ties consistently. Zero-weight resources may be dropped because removing their assignments preserves feasibility and reward.

A legal offline schedule may use each request at most once and must separate two uses of the same resource by at least $d$ slots. Its maximum reward is $\mathrm{OPT}$. Write $x_{it}\ge0$ for a fractional allocation supported on $i\in N_t$, and put $x_{it}=0$ outside the input. The full window relaxation maximizes $\sum_{i,t}w_i x_{it}$ subject to
\begin{equation}\label{eq:window-lp}
 \sum_i x_{it}\le1\quad\text{for every }t,
 \qquad
 \sum_{u=s}^{s+d-1}x_{iu}\le1\quad\text{for every }i,s.
\end{equation}
Its value is $\mathrm{LP}$. The second family is equivalently the backward-window constraint $\sum_{u\in W_t}x_{iu}\le1$ with
\[
 W_t=\{\max\{1,t-d+1\},\ldots,t\}.
\]
Each full forward window ends at a backward-window endpoint, and truncated boundary windows follow by nonnegativity after extending $x$ by zero. Every legal offline schedule gives a feasible allocation, so $\mathrm{LP}\ge\mathrm{OPT}$. At $d=2$ the LP is integral \cite[Theorem~4]{madarasi}. For $d\ge3$ the two values can differ, as in the example cited above.

\begin{theorem}[The uniform window-LP guarantee]\label{thm:main}
For every finite resource set $I$, known nonnegative weights $w$ and integer $d\ge1$, Algorithm~\ref{alg:continuous} is one online policy that, on every fixed finite menu sequence, satisfies
\[
  \mathbb E[\mathrm{ALG}]\ge (1-1/e)\sum_{i,t}w_i x_{it}
  \qquad\text{for every feasible allocation }x\text{ in \eqref{eq:window-lp}}.
\]
Its decisions use the known weights, the free current menu, its own past choices and winning bids, and the common clock. Their joint law on any revealed prefix is the same for every continuation and every horizon containing that prefix.
\end{theorem}

Write $c=1-1/e$. The order of quantifiers fixes the policy after $(I,w,d)$ and before the menu sequence or its length. Taking an LP optimum gives $\mathbb E[\mathrm{ALG}]\ge c\mathrm{LP}\ge c\mathrm{OPT}$. The upper bound over all durations is the classical ordinary-matching observation in Section~\ref{sec:upper}. Section~\ref{sec:fixed-clock} defines the optimal ratio at a fixed duration and establishes its distinct behavior.

\begin{table}[t]
\caption{Information used online. All policies know the resource weights and their own previous assignments.}\label{tab:information}
\small
\begin{tabularx}{\linewidth}{@{}p{.33\linewidth}X@{}}
\toprule
Policy or result & Information available before each decision\\
\midrule
Delayed restoration, including its strict gain & Free current menu and the integer clock with known $d$. Winning bids are computed by the policy. The LP allocation and width parameters enter only the proof.\\
Completion-based FIFO implementation (Appendix~\ref{sec:calendar}) & Free current menu and the identifiers of all own completed assignments, including completions outside the current menu.\\
Optimal duration-two policy & Complete current menu, including busy resources, and known $d=2$. It computes its offline prefix value from the revealed menus.\\
\bottomrule
\end{tabularx}
\end{table}

\begin{center}
\begin{minipage}{\linewidth}
\small
\textbf{Notation.}\par\smallskip
\begin{tabularx}{\linewidth}{@{}lX@{}}
\toprule
Symbol & Meaning\\
\midrule
$A_i$, $V_i(z)$ & Active rank region and bid at rank $z$.\\
$\mathcal H_r$, $\mathcal F_{it}$ & Winners and prices through the boundary, and the enlarged information that also reveals the competitor paths for $(i,t)$.\\
$Z_i(t)$, $b$ & Terminal potential rank and terminal price in the run with $i$ deleted.\\
$\beta_t$, $\alpha_{is}$ & Request payment and credit for a use of resource $i$ at slot $s$.\\
$\theta$, $y_{it}$ & Fraction of bid shifted to resource credit and comparator mass charging that credit.\\
$\Lambda(x)$, $H$ & Effective competitor mass and maximum width in the strict-gain proof.\\
$\lambda_i$ & Offline prefix loss on forbidding $i$ as the last action at duration two.\\
$c$, $\rho_{\mathcal B}(d)$ & $1-1/e$ and the best uniform guarantee at integer duration $d$.\\
\bottomrule
\end{tabularx}
\end{minipage}
\end{center}

The following weighted window accounting is the primal form of the approximate dual certificate in Delong et al.\ \cite[Section~2, Lemma~1]{delong}.
\begin{lemma}[Aggregation over a feasible window allocation]\label{lem:aggregation}
Suppose nonnegative payments $\beta_t,\alpha_{is}$ on a fixed input sum to its realized reward. For every feasible fractional allocation $x$, define
\[
 T(x)=\sum_{i,t}x_{it}\mathbb E\left[\beta_t+\sum_{s\in W_t}\alpha_{is}\right].
\]
Then $T(x)\le\mathbb E[\mathrm{ALG}]$. In particular, a weighted bound $T(x)\ge\rho(x)\sum_{i,t}w_i x_{it}$ implies the corresponding reward bound for that $x$. If every eligible edge has expected charge at least $\rho w_i$, the expected reward is at least $\rho\mathrm{LP}$.
\end{lemma}
\emph{Proof.}
Rearrange the finite sum and use both LP packing constraints,
\begin{align*}
 T(x)
 &=\sum_t\mathbb E[\beta_t]\sum_i x_{it}
   +\sum_{i,s}\mathbb E[\alpha_{is}]\sum_{u=s}^{s+d-1}x_{iu}\\
 &\le\mathbb E\left[\sum_t\beta_t+\sum_{i,s}\alpha_{is}\right].
\end{align*}
The request coefficient and the resource-window coefficient are each at most one. All payments belong to the same actual-run law even when each edge uses its own coupling. The weighted and uniform implications follow by taking the given lower bound on $T(x)$, then an LP optimum in the uniform case. Under the FIFO rule of Proposition~\ref{prop:calendar}, replace the resource coefficient by $\sum_{t:s\in W_t}x_{it}$; the last $u$ with $s\in W_u$ satisfies $[s,u]\subseteq W_u$, which bounds this coefficient by one.\hfill$\square$

\section{Delayed restoration}\label{sec:continuous}
A free resource $i$ holds a finite union $A_i$ of active intervals in $(0,1]$ and a calendar of disjoint regions due for restoration. Initially $A_i=(0,1]$ and the calendar is empty. Draw a rank on an active set $A$ using the cumulative distribution
\[
 F_A(z)=\exp\!\left(-\int_{A\cap(z,1]}\frac{dv}{v}\right),\qquad 0<z\le1,
 \qquad V_i(z)=w_i(1-e^{-z}).
\]
Here $V_i(z)$ is the bid of resource $i$ at rank $z$. The distribution $F_A$ is the law of the largest point of a Poisson process with intensity $dz/z$ on $A$, with maximum zero for an empty realization. The event that this maximum is at most $z$ is the event that there is no point in $A\cap(z,1]$. Algorithm~\ref{alg:continuous} makes fresh independent draws at each decision. Every realized active region contains $(0,\varepsilon]$ for some $\varepsilon>0$: finitely many strictly positive winning prices remove only ranks bounded away from zero. Thus its bids are positive almost surely. The recorded observation at an allocation is the selected resource and its numerical bid.

\begin{algorithm}[ht]
\caption{Delayed restoration at request $t$}\label{alg:continuous}
\begin{algorithmic}[1]
\State Reset each physically returning resource to $(0,1]$ with an empty calendar.
\State For each other free resource, restore all due regions to its active set and remove those calendar entries.
\State Form $E_t=\{i\in N_t:i\text{ is free and }w_i>0\}$.
\If{$E_t\ne\varnothing$}
    \State Independently draw $U_i\sim{\rm Unif}(0,1)$ and set $Z_i=F_{A_i}^{-1}(U_i)$ for each $i\in E_t$.
    \State Assign a maximizer $k$ of $V_i(Z_i)$ in the fixed tie order, record $b=V_k(Z_k)$, and set its physical return to $t+d$.
    \State Discard $k$'s rank state. For each $i\in E_t\setminus\{k\}$, form $S_{it}=A_i\cap\{z:V_i(z)>b\}$.
    \State Remove $S_{it}$ from $A_i$ and schedule every nonempty $S_{it}$ for restoration at $t+d$.
\EndIf
\end{algorithmic}
\end{algorithm}

\paragraph{Sampling the active region.}
List its intervals $(l,r]$ in descending order and initialize $P=1$. On one such interval the CDF equals $Pz/r$ and is constant in the following gap. If a uniform draw $U$ exceeds $Pl/r$, return $rU/P$. Otherwise continue with $P$ replaced by $Pl/r$. Endpoint ties have probability zero. This computes the generalized inverse of $F_A$ by a single interval scan. Section~\ref{sec:product} gives an equivalent way to generate the same winners and prices. It lets us compare two runs even when their earlier decisions differ.

\subsection{A persistent representation}\label{sec:product}
The proof uses a representation in which a losing resource retains its private rank. We first show that this representation has the observation law of Algorithm~\ref{alg:continuous}. We then construct its paths over a single window and compare the paths with and without a target resource.

For a fixed input, let $\mathcal H_s$ be the sigma-field generated by the winning resource and its numerical bid through slot $s$, together with a rejection symbol for an unmatched request. The input menus are fixed. The active regions, restoration calendars and physical release times are measurable functions of these observations. The retained ranks are hidden from $\mathcal H_s$.

For a finite union of intervals $B\subseteq(0,1]$, let $M(B)$ have cumulative distribution
\[
 \Pr\{M(B)\le z\}=F_B(z)
 =\exp\left(-\int_{B\cap(z,1]}\frac{dv}{v}\right),\qquad 0<z\le1.
\]
Any remaining mass is at zero, so $M(\varnothing)=0$. A region containing $(0,\varepsilon]$ has no atom at zero. For disjoint regions $B_1,\ldots,B_q$ and independent regional draws,
\begin{equation}\label{eq:max-union}
 \max_{1\le a\le q}M(B_a)\ \stackrel{\mathrm{law}}{=}\
 M\left(\bigcup_{a=1}^q B_a\right).
\end{equation}
This follows by multiplying the cumulative distributions. For a cutoff $z_\star$ with $F_A(z_\star)>0$,
\begin{equation}\label{eq:truncated-rank}
 \Pr\{M(A)\le z\mid M(A)\le z_\star\}
 =\frac{F_A(z)}{F_A(z_\star)}
 =\exp\left(-\int_{A\cap(z,z_\star]}\frac{dv}{v}\right),\quad 0<z<z_\star.
\end{equation}
Thus truncation gives $M(A\cap(0,z_\star])$, including when $z_\star$ lies in a gap of $A$.

The persistent process has state $(A_i,\mathcal D_i,R_i,Z_i)_{i\in I_+}$, where $I_+=\{i:w_i>0\}$, $\mathcal D_i$ is a finite calendar of pending regions and their deadlines, $R_i$ is the physical release time and $Z_i$ is the retained rank of a free resource. At a physical return, set $A_i=(0,1]$, clear $\mathcal D_i$ and draw $Z_i$ uniformly. When a pending region $D$ returns to a free resource, replace $Z_i$ by $\max\{Z_i,M(D)\}$ using a new independent draw. A losing resource retains its rank and undergoes the clipping in Algorithm~\ref{alg:continuous}. A selected resource discards its rank and calendar until return. These rules, the clock and fresh restoration draws define a Markov transition on the full state. Finite interval lists give a standard Borel state space. Appendix~\ref{app:window-measurability} supplies the state encoding, sampler measurability and conditional kernels used below.

\begin{lemma}[Conditional ranks after a decision]\label{lem:conditional}
Suppose that before a decision the free positive-weight resources have conditionally independent ranks with laws $M(A_i)$, given the recorded observations. After the decision, their surviving ranks, conditional on the new observations, have product law
\[
 \bigotimes_{i\text{ free},\ w_i>0}M(A_i'),
\]
where $A_i'$ are the updated active regions.
\end{lemma}
\emph{Proof.}
Let $E$ be the eligible free positive-weight resources. If $E$ is empty the observation gives no rank information. Otherwise the conditional bids are positive and atomless, because each active region contains an interval next to zero. The winner is unique almost surely.

Write $\mu_i$ for the predecision rank law and $\nu_k$ for the bid law of a prospective winner $k$. For measurable sets $C_i$ of the surviving ranks, the joint kernel of these ranks and the observation $(k,b)$ is
\begin{equation}\label{eq:winner-factor}
 \nu_k(db)
 \prod_{i\in E\setminus\{k\}}\mu_i(C_i\cap\{z:V_i(z)\le b\})
 \prod_{\substack{i\in I_+\text{ free}\\i\notin E}}\mu_i(C_i).
\end{equation}
The observation marginal is $\nu_k(db)\prod_{j\in E\setminus\{k\}}F_j^{\rm bid}(b)$, where $F_j^{\rm bid}(b)=\mu_j\{V_j\le b\}$. Consequently an eligible loser's posterior kernel is
\[
 Q_{j,b}(C)=\frac{\mu_j(C\cap\{V_j\le b\})}{F_j^{\rm bid}(b)}
 \quad\text{when }F_j^{\rm bid}(b)>0.
\]
At zero denominators choose any fixed probability law. Such prices have zero observation measure. Multiplying the kernels and integrating over a Borel price set recovers \eqref{eq:winner-factor}. A monotone class argument extends the identity from rank rectangles to all Borel sets. This defines conditioning at a numerical winning price through a regular conditional kernel.

Since $V_j$ is increasing, an eligible loser is truncated at $z_j^\star=-\log(1-b/w_j)$ when $b<V_j(1)$, and is unrestricted otherwise. Set $z_j^\star=1$ in the unrestricted case. Equation~\eqref{eq:truncated-rank} gives the updated law $M(A_j\cap(0,z_j^\star])$. Ineligible ranks keep their old laws, and the winner discards its rank. The posterior therefore factors as asserted. Positive winning prices give positive cutoffs. Finitely many such cutoffs preserve a nonempty interval next to zero.\hfill$\square$

\begin{lemma}[Conditional ranks after restoration]\label{lem:restore}
Condition on the observations after a decision and assume the product law of Lemma~\ref{lem:conditional}. After the due restorations and physical returns, before the next decision, the free positive-weight ranks have the product law on their enlarged active regions.
\end{lemma}
\emph{Proof.}
The recorded state specifies which regions return. For a resource remaining free, those regions are disjoint from its active region and from one another. Independent regional draws and \eqref{eq:max-union} give $M(A_i\cup D_i)$ after their restoration. A physically returning resource is reset with an independent uniform draw. The updates use independent draws for different resources, so the joint law remains a product.\hfill$\square$

\begin{corollary}[Equality of observation laws]\label{lem:fresh-law}
The persistent process and Algorithm~\ref{alg:continuous} have the same joint law of winners and numerical winning prices on every fixed finite input.
\end{corollary}
\emph{Proof.}
The initial ranks are independent uniforms. Lemmas~\ref{lem:conditional} and \ref{lem:restore} inductively give the product law before every decision. Its next-observation kernel equals that obtained by fresh independent inverse draws on the same active regions. Both processes apply the same deterministic calendar update to that observation. Induction proves the claim.\hfill$\square$

\subsection{Coupling a window to a deleted run}\label{sec:terminal}
Fix an eligible positive-weight edge $(i,t)$ and set $r=\max\{0,t-d\}$. The window is $W_t=\{r+1,\ldots,t\}$. We condition on the observations immediately after slot $r$, using the initial state if $r=0$. A deadline created by a slot $u\le r$ expires by $t$. Every assignment or suspension created at $s\in W_t$ has deadline $s+d>t$.

\paragraph{Boundary data and potential paths.}
Fix a boundary history $\eta$ in the probability-one set where the preceding conditional laws hold. For each $j\in I_+$, independently sample a Poisson process $\Pi_j$ of intensity $dz/z$ on $(0,1]$. It can be constructed by exponentiating the negatives of successive arrival times of a unit-rate Poisson process. The maximum in a region $B$ has law $M(B)$, and restrictions to disjoint regions are independent. Let $P_\eta$ be their product law.

For a resource free at the boundary, let $A_j^r$ be its active region and let $(D_{jm},e_{jm})$ be its disjoint pending regions and deadlines. These regions partition $(0,1]$. Put
\[
 B_j(s)=A_j^r\cup\bigcup_{m:e_{jm}\le s}D_{jm},
 \qquad Z_j(s)=\max(\Pi_j\cap B_j(s)),\qquad \tau_j=r+1.
\]
For a resource busy at the boundary, let $\tau_j\le t$ be its recorded release slot and put $B_j(s)=(0,1]$ and $Z_j(s)=\max\Pi_j$ for $s\ge\tau_j$. It cannot bid before $\tau_j$. Each potential path is nondecreasing after activation and continues after its resource is used in the window.

For $S=I_+$ or $S=I_+\setminus\{i\}$ define a deterministic map $\Phi_S(\eta,\Pi)$ as follows. Start with unused set $L_{r+1}^S=S$. At slot $s$, the candidates are
\[
 E_s^S=L_s^S\cap N_s\cap\{j:\tau_j\le s\}.
\]
Choose the largest $V_j(Z_j(s))$ in this set using the fixed tie order, record the winner and price, and remove only that resource from $L_s^S$. An empty candidate set records rejection at price zero. The future unused set tracks use since the boundary. It includes resources busy at the boundary, whose activation is enforced separately. Each map is measurable by the interval and Poisson evaluations in Appendix~\ref{app:window-measurability}.

\begin{lemma}[The revealed-point invariant]\label{lem:window-invariant}
On the point-cloud space, construct a version of the persistent process whose retained rank for a boundary-free resource $j$ is $\max(\Pi_j\cap A_j^r)$, whose draw on restoration of $D_{jm}$ is $\max(\Pi_j\cap D_{jm})$, and whose rank on a boundary-busy resource's return is $\max\Pi_j$. Before every decision and for each activated resource $j$ not yet used in $W_t$, its actual active region has the form
\[
 A_j(s)=B_j(s)\setminus Q_j(s),\qquad
 \Pi_j\cap Q_j(s)=\varnothing,
\]
where $Q_j(s)$ is the union of its suspensions created inside the window. Its retained rank is therefore $Z_j(s)$. The actual and $\Phi_{I_+}$ observations agree throughout the window.
\end{lemma}
\emph{Proof.}
Reveal $\Pi_j$ first on $A_j^r$ for a boundary-free resource and subsequently on $D_{jm}$ exactly at $e_{jm}$. For a boundary-busy resource, reveal its cloud at $\tau_j$. This schedule depends only on $\eta$. The boundary product law gives the prescribed initial conditional ranks. Independent increments give each newly revealed restriction its original Poisson law conditional on the restrictions revealed earlier. Each restoration draw thus has the required regional law and is independent of the previous full state and of the other restoration draws. Given $\eta$, this construction has the Markov transition kernels of the persistent process throughout the window.

At a resource's activation, $Q_j$ is empty and its retained rank is the prescribed regional maximum. Suppose the invariant holds before slot $s$. When an old pending region is restored, it is disjoint from every previously revealed region for that resource. In particular, it is disjoint from all suspensions created inside the window so far. Taking the maximum with its newly revealed points makes the retained rank the maximum on the enlarged $B_j(s)$.

If $j$ loses at price $b_s$, its retained bid is at most $b_s$. The new suspension is
\[
 S_{js}=A_j(s)\cap\{z:V_j(z)>b_s\}.
\]
Every revealed point lies at or below the retained maximum unless it is already in $Q_j(s)$, which has no point. Since $V_j$ is increasing, $\Pi_j\cap S_{js}=\varnothing$. Moreover $S_{js}\subseteq B_j(s)$, so it is disjoint from every old pending region that will be revealed later. Its deadline $s+d$ exceeds $t$. Replacing $Q_j(s)$ by $Q_j(s)\cup S_{js}$ therefore preserves both displayed statements through the remaining window, and the retained maximum is unchanged.

A selected resource is removed from the unused set and remains busy past $t$, so it requires no further invariant. These updates preserve equality of each candidate bid, and hence of the winner and price, at every slot. Induction gives the assertions.\hfill$\square$

\begin{lemma}[Actual marginal and terminal ranks]\label{lem:terminal}
For almost every boundary history $\eta$, a version of the conditional observation law of Algorithm~\ref{alg:continuous} is
\[
 \Pr\{(O_{r+1},\ldots,O_t)\in C\mid\mathcal H_r\}(\eta)
 =P_\eta\{\Pi:\Phi_{I_+}(\eta,\Pi)\in C\}
\]
for every Borel window-history set $C$. A resource used at $s\in W_t$ uses rank $Z_j(s)\le Z_j(t)$. The terminal ranks $Z_j(t)$ are independent uniforms, and the terminal price $b$ in $\Phi_{I_+\setminus\{i\}}$ is independent of $Z_i(t)$.
\end{lemma}
\emph{Proof.}
Lemma~\ref{lem:window-invariant} identifies the persistent run with $\Phi_{I_+}$. The two observation processes have the same joint distribution by Corollary~\ref{lem:fresh-law}. Given $\mathcal H_r$, the persistent ranks have the boundary product law of Lemma~\ref{lem:conditional}, and the construction in Lemma~\ref{lem:window-invariant} uses its conditional transition kernels. The two window laws given $\mathcal H_r$ therefore agree almost everywhere, which gives the displayed identity. For a boundary-free resource, all old deadlines have expired by $t$, so $B_j(t)=(0,1]$. A boundary-busy resource has returned by $t$ and uses the same full cloud. Thus every $Z_j(t)$ has CDF $F_{(0,1]}(z)=z$. The independent clouds give independence across resources. The deletion map uses only clouds other than $\Pi_i$, so its terminal price is independent of $Z_i(t)$. Monotonicity of $B_j(s)$ gives the comparison of ranks at use and at $t$.\hfill$\square$

\begin{lemma}[Deterministic deletion comparison]\label{lem:deletion}
For the two maps on any common realization, the actual terminal price $\beta_t$ is at least the deleted-run price $b$. If $V_i(Z_i(t))>b$, the actual map uses $i$ in $W_t$.
\end{lemma}
\emph{Proof.}
Before the first decision the actual unused non-target set equals the deletion set. Assume the former contains the latter before slot $s$. Both maps use the same activation, menu and priority tests on their common candidates. If the actual winner belongs to the deletion set, it is also the deletion winner. If the actual winner lies outside that set, its removal leaves all deletion candidates in the actual unused set; removing the deletion winner can only shrink the deletion set. Actual rejection implies deletion rejection. The inclusion holds before every decision, and it gives $\beta_t\ge b$.

Until the first actual use of $i$, the unused non-target sets are equal. When the actual choice is not $i$, both maps therefore make the same choice. If $i$ is unused before $t$, it is active and eligible at $t$ and faces exactly the deleted-run competitors. Its bid strictly above $b$ then wins.\hfill$\square$

\begin{theorem}[The edgewise window coupling]\label{thm:window-coupling}
For every positive-weight eligible edge $(i,t)$ and almost every boundary history $\eta$, the product space above supports the actual map $\Phi_{I_+}$, the deleted map $\Phi_{I_+\setminus\{i\}}$ and $Z_i(t)$ such that
\begin{enumerate}
\item $\Phi_{I_+}$ has Algorithm~\ref{alg:continuous}'s conditional observation law given $\eta$;
\item $Z_i(t)$ is uniform on $(0,1)$ and independent of the entire deleted map;
\item the actual terminal price satisfies $\beta_t\ge b$;
\item if $V_i(Z_i(t))>b$, the actual map uses $i$ at a slot $s\in W_t$ with $Z_i(s)\le Z_i(t)$.
\end{enumerate}
\end{theorem}
\emph{Proof.}
The first two assertions follow from Lemma~\ref{lem:terminal}, since the entire deleted map is a function of the other clouds. Lemma~\ref{lem:deletion} gives the last two assertions, together with the rank comparison in Lemma~\ref{lem:terminal}.\hfill$\square$

The construction may depend on the target edge. Each actual marginal is the same algorithmic law, so the expectations of its payments can be summed in Lemma~\ref{lem:aggregation}.

\begin{proposition}[Continuous edge charge]\label{prop:continuous-edge}
At a match of $j$ in slot $s$, put $\beta_s$ equal to its winning bid and $\alpha_{js}=w_j-\beta_s$. Set all other resource payments, and the price at an unmatched slot, to zero. Every eligible edge $(i,t)$ satisfies
\[
 \mathbb E\left[\beta_t+\sum_{s\in W_t}\alpha_{is}\right]
 \ge(1-e^{-1})w_i.
\]
\end{proposition}
\emph{Proof.}
Nonnegative payments cover a zero-weight target. For $w_i>0$, the resource is used at most once in $W_t$. Theorem~\ref{thm:window-coupling} gives, on each coupled realization,
\[
 \beta_t+\sum_{s\in W_t}\alpha_{is}
 \ge b+w_i e^{-Z_i(t)}
       {\bf1}\{w_i(1-e^{-Z_i(t)})>b\}.
\]
Condition on the boundary and all competitor clouds. They determine $b$, while $Z_i(t)$ is still uniform. If $b\ge cw_i$, where $c=1-e^{-1}$, the first term suffices. Otherwise set $z_\star=-\log(1-b/w_i)\in[0,1)$. The expected right side is
\[
 b+w_i\int_{z_\star}^1e^{-z}\,dz
 =b+w_i(1-b/w_i-e^{-1})=cw_i.
\]
Integrating over the boundary proves the proposition. Lemma~\ref{lem:aggregation} then proves Theorem~\ref{thm:main}'s lower bound.\hfill$\square$

\paragraph{Why earlier uses must be counted.}
To see why Proposition~\ref{prop:continuous-edge} counts a use of $i$ anywhere in the
window, consider the target $(i,3)$ with $d=2$ and $W_3=\{2,3\}$. Take unit weights,
$N_1=\{C,D\}$ and $N_2=N_3=\{i,B,C\}$. In the persistent representation
let the initial bids of $D,i,B,C$ be $0.60,0.50,0.40,0.20$, respectively.
Resource $C$ loses in slots 1 and 2. The two exclusions are due for
restoration before the decisions in slots 3 and 4, respectively.
The region restored to $C$ at slot 3 lies above a positive cutoff, so it contains no Poisson point with positive probability. On that event, $C$ still bids $0.20$.
The two runs use the same potential bids and first differ at slot 2.
\begin{center}
\begin{tabular}{@{}lll@{}}
\toprule
Run & Slot 2 & Slot 3\\
\midrule
Actual & $i$ wins at $0.50$ & $B$ is free and wins at $0.40$\\
With $i$ deleted & $B$ wins at $0.40$ & $B$ is busy and $C$ wins at $0.20$\\
\bottomrule
\end{tabular}
\end{center}
Thus $\beta_3=0.40$ and $b=0.20$. The actual use of $i$ that supplies its resource credit occurred one slot before the target request.

\paragraph{Why the suspension lasts exactly $d$.}
Both timing conditions in the window construction are needed. Suppose the suspension length were an integer $q>d$. A loss in boundary slot $r=t-d\ge1$ to an equal-weight rank $a\in(0,1)$ leaves the losing rank uniform on $(0,a)$ given that observation. Its removed region $(a,1]$ expires at $r+q>t$. If it is not offered again before $t$, its terminal potential remains supported below $a$ and is not uniform. Suppose instead $1\le q<d$ and a loss occurs at $s=t-q>r$. Given a cutoff $a$, the currently revealed cloud has no point above $a$. At its restoration at $t$, the persistent process draws a new $M((a,1])$, which exceeds $a$ with probability $1-a$. A cloud fixed at the boundary and already conditioned by that loss supplies no such point. Thus the reveal schedule used in Lemma~\ref{lem:window-invariant} no longer represents this modified process. These examples isolate the two identities used in this proof; they do not assert a competitive ratio for a modified timer.

\paragraph{Sampling and arithmetic complexity.}
For the example $A=(0,1/4]\cup(1/2,1]$, the sampler's CDF is
\[
 F_A(z)=
 \begin{cases}
 2z,&0<z\le1/4,\\
 1/2,&1/4<z\le1/2,\\
 z,&1/2<z\le1.
 \end{cases}
\]
A descending interval scan implements its generalized inverse. At each loss, clipping at one cutoff introduces at most one new endpoint. Expiry changes the active status of existing pieces; physical return removes all old endpoints. Hence a resource has at most $O(t)$ pieces by slot $t$. The number of active intervals can grow even at $d=2$. To see this, let one resource lose in slots $1,\ldots,k$ to distinct unit-weight rivals at strictly decreasing rank cutoffs $1>a_1>\cdots>a_k>0$, with all weights one. Put $S_0=\varnothing$. Just before decision $s$, the only unexpired suspension is $S_{s-1}$, so the newly suspended set is $S_s=(a_s,1]\setminus S_{s-1}$. For even $k$, $S_k$ consists of the alternating intervals $(a_k,a_{k-1}],\ldots,(a_2,a_1]$. Thus the available region at $k+1$ has $k/2+1$ components. Every finite sequence of these strict comparisons occurs with positive probability if the rival ranks lie in prescribed disjoint decreasing open intervals bounded away from zero and the target draws below them at each decision.

Store the endpoints and their readiness times. A slot at which the resource is ineligible creates no new suspension. Comparing its timestamps with the current slot therefore applies expiries lazily when the resource is next offered. A descending scan samples from the active intervals. Once the winning price is known, a second scan applies the cutoff to each losing resource. This gives $O(t)$ storage per resource and $O(|E_t|t)$ arithmetic operations at a decision with free positive-weight offers $E_t$, plus the scan of the revealed menu. These are bounds in a real-arithmetic model, which admits exact uniform draws and comparisons of the displayed bids. For an alternative with $m$ discrete rank levels, Appendix~\ref{app:formal-statements} gives the sampling rule and the formal guarantee. The real-arithmetic bound above counts neither random bits nor the bit operations required for price comparisons.

\subsection{The ordinary-matching upper bound}\label{sec:upper}
The nonreusable case gives an upper bound on any guarantee uniform over $d$ \cite{kvv}. Take $n$ unit-weight resources, $d=T=n$, and offer successive suffixes of a uniformly random permutation. The offline schedule serves the resource that expires from the next suffix and obtains $n$.

Write $x_t$ for the unconditional probability that an online algorithm serves request $t$. Conditional on the revealed suffixes and the algorithm's private history, the last permutation resource is uniform among the $n-t+1$ surviving names. This symmetry holds because the past menus reveal only the expired prefix, and the private decisions do not reveal the remaining order. The final resource is served at most once, so
\[
 \sum_{t=1}^n\frac{x_t}{n-t+1}\le1.
\]
At fixed $S=\sum_t x_t$, this sum is minimized by filling the earliest requests first, since their coefficients are smallest and $0\le x_t\le1$. The minimum is at least $\int_0^S(n+1-u)^{-1}\,du$. Hence $S\le c(n+1)$. Averaging yields one fixed permutation with this upper bound. Letting $n$ grow excludes every uniform ratio greater than $c$ against the integer optimum. This is the classical upper-bound construction, and its duration grows with its size.

\begin{corollary}[Fixed duration in real time]\label{cor:real}
For every prescribed real $D>0$ and arbitrary nondecreasing arrival times processed in a fixed input order, delayed restoration attains $1-1/e$ against the occupancy LP and the integer optimum. The ratio is sharp when arrival density is unrestricted. Both claims hold when a use at time $\tau$ becomes reusable at time $\tau+D$, and when reuse requires a strictly later time.
\end{corollary}
\emph{Proof.}
Let $\tau_t$ be the time of request $t$. Under immediate reuse at expiry, let $k_t$ be the last index with $\tau_s\le\tau_t-D$, or zero if none exists. Under the strict endpoint convention used by Delong et al.\ \cite[Appendix~A]{delong}, use $\tau_s<\tau_t-D$ instead. The cutoffs are nondecreasing, including at ties. Process simultaneous requests in their fixed input order. At a given timestamp the last occupancy constraint includes all tied requests, and the earlier constraints use successive prefixes. The indices $k_t+1,\ldots,t$ describe exactly the backward occupancy window, respectively $(\tau_t-D,\tau_t]$ or $[\tau_t-D,\tau_t]$. The FIFO transfer in Proposition~\ref{prop:calendar} gives the lower bound. For sharpness, place the permutation instance strictly inside one interval of length $D$.\hfill$\square$

\section{Strict improvement at fixed integer duration}\label{sec:gain}
The uniform guarantee uses only part of the payment available on a target edge. An earlier assignment can contribute resource credit even when its terminal potential bid is below the deletion price. A resource first assigned at the target request can pay more than that price. These two sources of surplus give a strict improvement at every fixed integer duration. The proof first identifies the prices at which neither source gives enough surplus, then pays for those exceptional comparisons using resource credit left over in the aggregate charge.

The argument also covers the common FIFO calendar of Proposition~\ref{prop:calendar}. In that setting define $W_t=\{k_t+1,\ldots,t\}$, where $k_t$ is the last expired origin at arrival $t$. For the integer clock, $k_t=\max\{0,t-d\}$. All fractional assignments below are feasible for the corresponding occupancy relaxation.

Write $c=1-e^{-1}$. For an eligible edge $(i,t)$ with $w_i>0$ define
\[
 \kappa_{it}=\sum_{j\in N_t\setminus\{i\}:\, c w_j\ge w_i/5}
            \min\{1,w_i/w_j\}.
\]
Here $N_t$ is the complete menu of the input on which Algorithm~\ref{alg:continuous} is run. Only a competitor whose largest bid reaches $w_i/5$ enters the sum. A heavier competitor has a smaller coefficient because its bid has lower density in units of $w_i$. For a feasible allocation $x$ of value $M=\sum_{i,t}w_i x_{it}>0$, put
\[
 \kappa(x)=\frac1M\sum_{t,\,i\in N_t:\,w_i>0} w_i x_{it}\kappa_{it}.
\]

Fix a feasible allocation $x$ of positive value. We choose the payment split $0<\theta\le2^{-16}$ as a function of $x$ in the proof. The decisions of Algorithm~\ref{alg:continuous} depend on neither $x$ nor $\theta$. Write $\beta_t^\theta=(1-\theta)\beta_t$ for the request payment. A use of $i$ at $t$ receives resource credit $\alpha_{it}^\theta=w_i-\beta_t^\theta$; every other resource receives zero credit there. Both payments are nonnegative and sum to the realized reward.

Fix a positive target edge $(i,t)$ and express comparison prices in units of $w_i$. Condition on the winners and prices at the window boundary. Let $b$ be the terminal price with $i$ deleted, let $P_i$ be $i$'s terminal potential bid, and let $\chi_{it}=w_i^{-1}(\beta_t^\theta+\sum_{s\in W_t}\alpha_{is}^\theta)$ be the normalized edge charge. Write $\mathcal F_{it}$ for the boundary history together with all competitor paths. It determines $b$. By Lemma~\ref{lem:terminal}, $P_i$ is independent of $\mathcal F_{it}$ and has CDF $-\log(1-p)$ on $[0,c]$.

\begin{lemma}[Baseline charge]\label{lem:gain-baseline}
On every coupled realization,
\[
 \chi_{it}\ge Y_\theta=(1-\theta)b+[1-(1-\theta)P_i]\mathbf1\{P_i>b\}.
\]
Its conditional mean is
\[
 \psi_\theta(b)=\mathbb E[Y_\theta\mid\mathcal F_{it}]=
 \begin{cases}
 (1-\theta)c+\theta[1+\log(1-b)],&0\le b\le c,\\
 (1-\theta)b,&b>c.
 \end{cases}
\]
Put $\ell=1/5$. If $b\le\ell$ or $b\ge c+2\theta$, then $\psi_\theta(b)\ge c+\theta/8$. For every $b\ge0$, $\psi_\theta(b)\ge c-2\theta/3$.
\end{lemma}
\emph{Proof.}
The deletion comparison gives the pointwise bound exactly as in Proposition~\ref{prop:continuous-edge}, with the scaled payments. At $\theta=0$ its integral is $c$ for $0\le b\le c$. Scaling that baseline by $1-\theta$ leaves an additional resource payment of mean $\theta\Pr(P_i>b)=\theta[1+\log(1-b)]$. For $b>c$ only the arrival term remains. The lower-price bound uses $1/e+\log(4/5)>1/8$. The upper-price bound follows from $(1-\theta)(c+2\theta)\ge c+\theta/8$. Finally $\psi_\theta(b)\ge(1-\theta)c\ge c-2\theta/3$.\hfill$\square$

\begin{lemma}[A single high point]\label{lem:gain-one-point}
In normalized bid coordinates the target Poisson process has intensity
\[
 \iota(p)\,dp=\frac{dp}{(1-p)(-\log(1-p))},\qquad 0<p<c.
\]
Let $\mathcal J=[\ell,c]$ and let $\mathcal L_i$ record its points below $\ell$. The event of exactly one point in $\mathcal J$, with location in $dp$, has density
\[
 \exp\left(-\int_{\mathcal J}\iota(v)\,dv\right)\iota(p)
       =[-\log(1-\ell)]\iota(p)>1/2,
\]
independently of $\mathcal L_i$ and all competitor paths.
\end{lemma}
\emph{Proof.}
Apply the change of variable $p=1-e^{-z}$ to the rank intensity $dz/z$. Independent Poisson restrictions give the stated one-point measure and independence. For $0<p\le c$, write $v=1-p\in[1/e,1)$. The function $-v\log v$ is at most $1/e$, with equality at $v=1/e$. Consequently
\[
 [-\log(1-\ell)]\iota(p)
 \ge e[-\log(4/5)]>e/5>1/2.
\]
This gives the claimed lower bound uniformly over $\mathcal J$.\hfill$\square$

For each target price $p$, let $f_i(p)$ be the first slot when that coordinate is active under the boundary schedule. A resource free at the boundary activates its old active region immediately and its pending regions at their recorded deadlines; a resource busy at the boundary activates all coordinates on physical return. Thus $f_i$ is fixed at the boundary. Before the first use, the bid is the largest point whose activation date has arrived. A loss cannot remove any already present point above the winning price, since such a point would have won.

\begin{lemma}[Two deletions and early use]\label{lem:gain-early}
Fix a positive-weight competitor $j\in N_t\setminus\{i\}$. If $j$ wins at $t$ in the $i$-deleted run, removing $j$ as well preserves all winners and prices at $u\in W_t$, $u<t$. Let $v_u$ be those double-deletion prices, with rejection priced at zero. There is a measurable set $\mathcal U_j\subseteq\mathcal J$, determined by the boundary, $\mathcal L_i$ and the paths other than $i,j$, such that, on the one-high-point event with location $p$,
\[
 i\text{ is used before }t\quad\Longleftrightarrow\quad p\in\mathcal U_j
\]
almost surely whenever $j$ is the $i$-deleted terminal winner.
\end{lemma}
\emph{Proof.}
A terminal winner was not used earlier in this window. Deleting that unused resource leaves every earlier choice and price unchanged, by induction over the slots. If an eligible, activated low point of $i$ beats $v_u$ at some $u<t$, put $\mathcal U_j=\mathcal J$. Otherwise put
\[
 \mathcal U_j=\{p\in\mathcal J:\ \exists u\in W_t,\ u<t,\ i\in N_u,\ f_i(p)\le u,\ p>v_u\}.
\]
Until $i$ first wins, its competitors make exactly their deletion-run choices. The first active target point to beat an eligible deletion price therefore gives its first use. The set $\mathcal U_j$ can depend on the revealed low target points $\mathcal L_i$. It reads neither the point cloud of $j$ nor any high point of $i$. Conditional on these defining variables, a positive high target point ties a comparison price with probability zero. This proves the equivalence. The low-point test and the displayed finite union are measurable, as are the lengths of their sections.\hfill$\square$

\begin{lemma}[Exceptional prices]\label{lem:gain-exceptional}
Let $E_j$ be the set of $b\in\mathcal J$ satisfying both
\[
 |\mathcal U_j\cap[\ell,b)|<8\theta,\qquad
 |(\mathcal J\setminus\mathcal U_j)\cap(b,c]|<3\sqrt\theta.
\]
Then $|E_j|\le8\theta+3\sqrt\theta$. If $E_{it}^{\rm bad}$ is the event that the $i$-deleted winner is some $j$ and $b\in E_j\cup[c,c+2\theta)$, then
\begin{equation}\label{eq:bad-prob}
 \Pr(E_{it}^{\rm bad})\le9\kappa_{it}(10\theta+3\sqrt\theta).
\end{equation}
The probability bound holds conditional on almost every boundary history as well.
\end{lemma}
\emph{Proof.}
For $b_1<b_2$ in $E_j$, the portion of $(b_1,b_2)$ in $\mathcal U_j$ is counted at $b_2$, and its complement is counted at $b_1$. Their sum bounds $b_2-b_1$ by $8\theta+3\sqrt\theta$.

For each fixed positive-weight $j\in N_t\setminus\{i\}$, reveal the boundary, $\mathcal L_i$ and every competitor path except $j$'s. In particular, $\mathcal L_i$ is held fixed throughout this density integration. These variables determine $E_j$, independently of the cloud of $j$. With $r_j=w_j/w_i$, the full terminal bid $P_j$ has density $1/(r_j-b)$ on $(0,cr_j)$. For $b\ge\ell$, this density is at most $9\min\{1,1/r_j\}\mathbf1\{cr_j\ge\ell\}$, since on its support $r_j-b\ge r_j/e$ and $1/(r_j-b)\le(e-1)/b$. On the event that $j$ wins, $b=P_j$. Hence
\[
 \{j\text{ wins with }i\text{ deleted},\ b\in E_j\cup[c,c+2\theta)\}
 \subseteq\{P_j\in E_j\cup[c,c+2\theta)\}.
\]
The right side tests the full bid against a fixed set of length at most $10\theta+3\sqrt\theta$. Integrating its density before restricting to the winning event, and summing over terminal candidates $j\in N_t\setminus\{i\}$, proves \eqref{eq:bad-prob}. A rejection has $b=0$ and lies outside this event.\hfill$\square$

\begin{lemma}[Surplus outside the exceptional set]\label{lem:gain-surplus}
Fix $\mathcal F_{it}$ and $\mathcal L_i$. If $b\in\mathcal J$ has terminal winner $j$, then
\[
 \mathbb E[\chi_{it}-Y_\theta\mid\mathcal F_{it},\mathcal L_i]
       \ge\theta\mathbf1\{b\notin E_j\}.
\]
Consequently, after integrating the entire target cloud,
\[
 \mathbb E[\chi_{it}\mid\mathcal F_{it}]
       \ge \psi_\theta(b)+\theta\Pr(b\notin E_j\mid\mathcal F_{it}),\qquad b\in\mathcal J.
\]
\end{lemma}
\emph{Proof.}
On the one-high-point event the terminal target bid is $P_i=p$. Suppose $b\notin E_j$. If $|\mathcal U_j\cap[\ell,b)|\ge8\theta$, a sole high point $p$ in this set makes $i$ win early at a bid at most $p<b$. The baseline omits its entire resource credit, at least $1-(1-\theta)p\ge1/e$. Lemma~\ref{lem:gain-one-point} gives expected surplus at least $(1/2)(8\theta)/e>\theta$.

Otherwise $|(\mathcal J\setminus\mathcal U_j)\cap(b,c]|\ge3\sqrt\theta$. A sole high point $p$ in this set first wins at $t$ and adds $(1-\theta)(p-b)$. Among subsets of $(b,c]$ of length $m$, the integral of $p-b$ is smallest on $(b,b+m]$ and equals $m^2/2$. The surplus is therefore at least $(1-\theta)9\theta/4>\theta$. Every other realization has nonnegative surplus by Lemma~\ref{lem:gain-baseline}. This proves the first inequality, which integrates the high points. Average it over $\mathcal L_i$ and use $\mathbb E[Y_\theta\mid\mathcal F_{it}]=\psi_\theta(b)$ to get the second. The baseline is thus averaged over the entire target cloud even though $E_j$ depends on its low points.\hfill$\square$

\begin{lemma}[A bound in terms of competing bids]\label{lem:menu-gain}
For every fixed input and feasible fractional allocation $x$ of value $M>0$, delayed restoration earns at least $(c+2^{-21}/(1+\kappa(x))^2)M$ in expectation.
\end{lemma}
\emph{Proof.}
Set $\theta=[256(1+\kappa(x))]^{-2}$. On the interior price range $\mathcal J$, Lemma~\ref{lem:gain-surplus} adds $\theta$ off $E_j$ to a baseline of at least $c-2\theta/3$. Outside $\mathcal J$, Lemma~\ref{lem:gain-baseline} gives at least $c+\theta/8$ except on the upper strip already included in $E_{it}^{\rm bad}$, where the baseline is at least $c-2\theta/3$. Averaging over the target points gives
\[
 \mathbb E\chi_{it}\ge c+\theta/8-(19/24)\theta\Pr(E_{it}^{\rm bad}).
\]
Under the weights $w_ix_{it}/M$, Lemma~\ref{lem:gain-exceptional} bounds the mean exceptional probability by $9\kappa(x)(10\theta+3\sqrt\theta)\le90/256^2+27/256<1/9$. Since $1/8-(19/24)/9\ge1/32$, Lemma~\ref{lem:aggregation} gives expected reward at least $(c+\theta/32)M$.\hfill$\square$

\paragraph{Menus of bounded size.}
If each menu has at most $\Delta$ positive-weight resources, then $\kappa(x)\le\Delta-1$. Lemma~\ref{lem:menu-gain} therefore gives $(c+2^{-21}/\Delta^2)\mathrm{LP}$ for any duration. When no resource returns, the retained ranks stay fixed until their resource is assigned, so the same bound applies to vertex-weighted Ranking. The next argument replaces the number of competitors by the number that can carry substantial comparator mass in one window.

\subsection{Accounting for uncharged resource credit}\label{sec:untrimmed}
The aggregation in Lemma~\ref{lem:aggregation} can leave part of a resource payment uncharged. We first record that remainder under the same payment split as the target-edge estimates, then use it when a terminal winner has little mass in the comparator.

Fix feasible $x$ of value $M>0$. Put $\ell=1/5$, $y_{rt}=\sum_{u:\,t\in W_u}x_{ru}$ and $S_t=\{r\in N_t:w_r>0,\ y_{rt}\ge1/2\}$. The coefficient $y_{rt}\le1$ is the charged fraction of resource $r$'s credit at $t$. Let $\mathcal T$ be the requests offering at least two positive-weight resources and define
\[
 \nu_{it}=\mathbf1_{\{t\in\mathcal T\}}
       \sum_{r\in S_t:\,cw_r\ge\ell w_i}\min\{1,w_i/w_r\},
 \qquad
 \Lambda(x)=\max\left\{1,\frac1M\sum_{(i,t):\,w_ix_{it}>0}
                               w_ix_{it}\nu_{it}\right\}.
\]
The sum may include the target $i$. To state the width bound precisely, set
\[
 \Gamma_t=\{t\}\cup
 \{u>t:t\in W_u,\ \exists r\in N_t\cap N_u\text{ with }w_r>0\},
 \qquad H=\max_{1\le t\le T}|\Gamma_t|.
\]
Besides $t$, only requests sharing a positive-weight conflicting offer enter $\Gamma_t$. Set $H=1$ on an empty input, whose reward bound is immediate.

\begin{theorem}[Strict gain at fixed duration]\label{thm:untrimmed}
On the original menus, Algorithm~\ref{alg:continuous}, or its common-FIFO implementation in Proposition~\ref{prop:calendar}, earns at least
$(c+2^{-21}/\Lambda(x)^2)M$ in expectation. It also guarantees
\[
 \mathbb E[\mathrm{ALG}]
       \ge\left(c+\frac{2^{-21}}{H^2}\right)\mathrm{LP}.
\]
If every backward occupancy window of the common FIFO calendar contains at most $h_0$ requests, with $h_0\ge1$, then $H\le h_0$ and the bound also holds with $h_0$ in place of $H$. In particular, on the integer clock,
\[
 \mathbb E[\mathrm{ALG}]\ge\left(c+\frac{2^{-21}}{d^2}\right)\mathrm{LP}.
\]
The algorithm uses the original free eligible offers, its own choices, the winning bids and the common clock or the completion notifications of Proposition~\ref{prop:calendar}; neither $x$ nor a bound on $H$ is supplied to it.
\end{theorem}

The proof applies the preceding estimates to the same untrimmed execution. The allocation $x$ and the payment split below are used only in the analysis.

For $0<\theta\le2^{-16}$ use the payments $\beta_t^\theta,\alpha_{is}^\theta$ defined before Lemma~\ref{lem:gain-baseline}. Let $G_{\rm alg}=\mathbb E[\mathrm{ALG}]$, let $T_\theta(x)$ be their aggregate comparator charge, and write $U_\theta=G_{\rm alg}-T_\theta(x)$.

\begin{lemma}[Uncharged resource credit]\label{lem:uncharged-credit}
For $z_t=\sum_i x_{it}$,
\[
 U_\theta
   =\sum_t(1-z_t)\mathbb E\beta_t^\theta
       +\sum_{r,t}(1-y_{rt})\mathbb E\alpha_{rt}^\theta
   \ge\sum_t\mathbb E\big[(1-y_{J_t,t})\alpha_{J_t,t}^\theta\big].
\]
Here $J_t$ is the actual winner and the last summand is zero on rejection.
\end{lemma}
\emph{Proof.}
Expand $T_\theta(x)$ and reverse its finite sums. The request payment at $t$ has coefficient $z_t$, and the resource credit at $(r,t)$ has coefficient $y_{rt}$. Subtract from the total expected payment $G_{\rm alg}$. Both coefficients lie in $[0,1]$. Only $J_t$ receives resource credit at $t$, giving the inequality. All terms use the same $\theta$ and the actual policy law.\hfill$\square$

For each positive target edge $(i,t)$ measure the comparison prices $b_{\rm act},b$ and the edge charge $\chi_{it}$ in units of $w_i$. Keep $T_\theta$, $U_\theta$ and the individual payments $\alpha_{rt}^\theta$ in the original reward units. Set $\varepsilon_{\rm gap}=8\theta$ and $\zeta=19/24$. Let $b_{\rm act}$ and $b$ be the actual and $i$-deleted terminal prices and put $\mathcal E=E_{it}^{\rm bad}$. On $\mathcal E$ we have $b_{\rm act}\ge b\ge\ell$; let $r,j$ be the actual and deleted terminal winners. Partition this event into
\[
 \begin{aligned}
 \mathcal E_1&=\mathcal E\cap\{b_{\rm act}-b>\varepsilon_{\rm gap}\},\\
 \mathcal E_2&=\mathcal E\cap\{b_{\rm act}-b\le\varepsilon_{\rm gap},\ r\in S_t\},\\
 \mathcal E_3&=\mathcal E\cap\{b_{\rm act}-b\le\varepsilon_{\rm gap},\ r\notin S_t\}.
 \end{aligned}
\]
Each event is defined on its target-edge coupling.

\begin{lemma}[Payment for a large price gap]\label{lem:untrimmed-gap}
The normalized charge satisfies
\[
 \mathbb E\chi_{it}
       \ge c+\theta/8-\zeta\theta
                    \big[\Pr(\mathcal E_2)+\Pr(\mathcal E_3)\big].
\]
\end{lemma}
\emph{Proof.}
The nonnegative surplus $\chi_{it}-Y_\theta$ is at least $(1-\theta)(b_{\rm act}-b)$: the resource payment already covers the indicator term of $Y_\theta$. Both lower bounds apply to the same nonnegative surplus. Their convex combination assigns seven eighths to the estimate in Lemma~\ref{lem:gain-surplus} and one eighth to the price difference. This combination counts each part of the surplus once. For interior prices outside $E_j$, the first part adds at least $7\theta/8$, so $-2/3+7/8\ge1/8$. Lemma~\ref{lem:gain-baseline} treats the exterior prices. After averaging the low target points, as in Lemma~\ref{lem:gain-surplus}, we obtain
\[
 \mathbb E\chi_{it}
 \ge c+\theta/8-\zeta\theta\Pr(\mathcal E)
          +\frac{1-\theta}{8}\mathbb E(b_{\rm act}-b).
\]
On $\mathcal E_1$ the last payment exceeds $(1-\theta)\theta\ge \zeta\theta$. The price difference is nonnegative elsewhere. This pays the entire loss assigned to $\mathcal E_1$.\hfill$\square$

\begin{lemma}[A close winner with large charged fraction]\label{lem:untrimmed-density}
\[
       \Pr(\mathcal E_2)
           \le9\nu_{it}(18\theta+3\sqrt\theta).
\]
\end{lemma}
\emph{Proof.}
At $t\notin\mathcal T$ the deletion price is zero and $\mathcal E$ is empty. Otherwise first consider $r=j\in S_t$. Restrict the union in Lemma~\ref{lem:gain-exceptional} to $S_t$ and retain each density factor. Its bound is $9\nu_{it}(10\theta+3\sqrt\theta)$.

For $r\ne j$, Lemma~\ref{lem:deletion} puts $j$ among the actual free terminal candidates, with its potential bid $b$. Fix $r\in S_t$ and delete $r$ from the actual window run, obtaining terminal price $b^{(-r)}$. On the event that $r$ is the actual terminal winner, it was unused earlier in the window. Its deletion preserves that entire earlier prefix. Hence $b\le b^{(-r)}\le b_{\rm act}$ when $r\ne j$. The additional requirement $b_{\rm act}-b\le\varepsilon_{\rm gap}$ forces the full terminal bid of $r$ into
\[
       [b^{(-r)},b^{(-r)}+\varepsilon_{\rm gap}]\cap[\ell,\infty).
\]
Given the boundary and every path except $r$'s, this interval is fixed. Above $\ell$ the bid has density at most
$9\min\{1,w_i/w_r\}\mathbf1_{\{cw_r\ge\ell w_i\}}$,
by the density calculation in Lemma~\ref{lem:gain-exceptional}; the same calculation applies when $r=i$. Integrate before imposing that $r$ wins, then sum over $S_t$. This contributes at most $9\nu_{it}\varepsilon_{\rm gap}$. Adding the two cases proves the bound.\hfill$\square$

\begin{lemma}[A close winner with uncharged credit]\label{lem:untrimmed-credit}
\[
 \sum_{(i,t):\,w_ix_{it}>0}w_ix_{it}\Pr(\mathcal E_3)
       \le\frac{2c}{\ell(1-c)}U_\theta .
\]
\end{lemma}
\emph{Proof.}
On $\mathcal E_3$, the actual price in the original units is $w_i b_{\rm act}\ge \ell w_i$. The bid cap $cw_r$ therefore gives $w_r\ge\ell w_i/c$ and
$\alpha_{rt}^\theta\ge(1-c)w_r$. Since $r\notin S_t$, its uncharged fraction exceeds one half. Therefore
\[
 w_i\mathbf1_{\mathcal E_3}
 \le\frac{2c}{\ell(1-c)}
               (1-y_{J_t,t})\alpha_{J_t,t}^\theta .
\]
The right side is evaluated on the actual path and is zero on rejection. Its expectation is the same under every target coupling at $t$. Multiply by $x_{it}$, sum over $i$ using $\sum_i x_{it}\le1$, and sum over $t$. Lemma~\ref{lem:uncharged-credit} gives the claim.\hfill$\square$

\emph{Proof of Theorem~\ref{thm:untrimmed}.}
Take $\theta=(256\Lambda(x))^{-2}$. By Lemma~\ref{lem:untrimmed-density}, the value-weighted mean $p_{\rm exc}=M^{-1}\sum_{i,t}w_ix_{it}\Pr(\mathcal E_2)$ is at most
\[
 9\Lambda(x)(18\theta+3\sqrt\theta)
     \le162/256^2+27/256<1/9 .
\]
The gap and uncharged-credit lemmas imply
\[
 T_\theta(x)\ge[c+\theta(1/8-\zeta p_{\rm exc})]M
               -\frac{2\zeta\theta c}{\ell(1-c)}U_\theta .
\]
Add $U_\theta$ to both sides. Its coefficient on the right remains nonnegative because $2\zeta\theta c/[\ell(1-c)]<1$. Since $1/8-\zeta/9\ge1/32$, this gives $G_{\rm alg}\ge(c+\theta/32)M$.

For the width assertion, support on offered pairs gives
\[
 \sum_{r\in N_t:\,w_r>0}y_{rt}
       \le\sum_{u\in\Gamma_t}\sum_r x_{ru}\le H .
\]
For the integer clock, every member of $\Gamma_t$ lies in $\{t,\ldots,t+d-1\}$, and hence $H\le d$. Under a common FIFO calendar, $\Gamma_t$ lies in the occupancy window of its last member. Indeed, if $v=\max\Gamma_t$, then $t\in W_v$; every $u\in\Gamma_t$ satisfies $t\le u\le v$, so $u\in W_v$. Thus $H\le h_0$ whenever all backward windows have at most $h_0$ requests. Repeat the lemmas with
$S_t=\{r\in N_t:w_r>0,\ y_{rt}>H/(H+1)\}$
and $\theta=(256H)^{-2}$. The strict threshold gives $|S_t|\le H$, so $\nu_{it}\le H$. Outside $S_t$, the unused fraction is at least $1/(H+1)$; the coefficient in the credit lemma becomes $c(H+1)/[\ell(1-c)]$. Its product with $\zeta\theta$ is less than one for every $H\ge1$. The same aggregation yields $c+\theta/32=c+2^{-21}/H^2$. Take an LP optimum.\hfill$\square$

\paragraph{The size and order of the gain.}
Table~\ref{tab:gain-constants} records the constants used above. They were chosen to make every exceptional event payable by the available surplus. The resulting coefficient $2^{-21}$ is a certified lower bound. The proof does not optimize it.

\begin{table}[t]
\caption{Constants in the strict-gain argument.}\label{tab:gain-constants}
\centering\small
\begin{tabular}{@{}p{0.19\linewidth}p{0.31\linewidth}p{0.43\linewidth}@{}}
\toprule
Constant & Value or scale & Role in the proof\\
\midrule
$\ell$ & $1/5$ & Separates low deletion prices from the high-point region.\\
Early-use threshold & $8\theta$ & A set of this length supplies surplus at least $\theta$.\\
Overshoot threshold & $3\sqrt\theta$ & Its squared length pays for a terminal winning bid above the deletion price.\\
$\varepsilon_{\rm gap}$ & $8\theta$ & A larger actual--deleted price gap pays for its exceptional event.\\
$\zeta$ & $19/24$ & Difference between the good-charge coefficient $1/8$ and the baseline coefficient $-2/3$.\\
$\theta$ & $(256\Lambda(x))^{-2}$ & Makes the mean exceptional probability smaller than $1/9$.\\
Width choice & $(256H)^{-2}$ & Gives the uniform bound on the integer clock through $H\le d$.\\
\bottomrule
\end{tabular}
\end{table}

The inverse square enters through the overshoot estimate in Lemma~\ref{lem:gain-surplus}. A favorable set of prices of length $m$ contributes at least a constant times $m^2$. Obtaining surplus of order $\theta$ therefore uses a set of length of order $\sqrt\theta$. The density bound for the $H$ relevant competitors then costs order $H\sqrt\theta$ in exceptional probability. Keeping that probability below a fixed constant leads to $\theta$ of order $H^{-2}$. Thus the exponent in this guarantee comes from the present estimates. Its optimality for delayed restoration remains open. For the best policy at a fixed duration, the permutation upper bound in Section~\ref{sec:fixed-clock} allows an excess of order $1/d$; the lower and upper bounds leave a gap in the exponent.

\section{A fixed integer clock}\label{sec:fixed-clock}
Fix an integer duration $d$ and a benchmark $\mathcal B\in\{\mathrm{OPT},\mathrm{LP}\}$. A policy family may depend on the finite resource set $I$, its known nonnegative weight vector $w$, and $d$. For each such triple, it must use one online rule for every finite menu sequence and every horizon, with identical decision laws on common revealed prefixes. The value $\rho_{\mathcal B}(d)$ is the supremum of the numbers $r\in[0,1]$ for which
\[
\begin{gathered}
 \forall(I,w)\ \exists\pi_{I,w,d}\ \forall T,N_1,\ldots,N_T,\\
 \mathbb E[\operatorname{reward}_{\pi_{I,w,d}}]
 \ge r\mathcal B(I,w,d,N_1,\ldots,N_T).
\end{gathered}
\]
Policies in this definition receive each complete menu, including its busy resources. Delayed restoration attains its bound using only the free eligible resources, whereas the duration-two policy below uses the complete menus. The order of quantifiers permits dependence on known weights and excludes dependence on future menus or the termination time.

\begin{proposition}[Monotonicity in the integer duration]\label{prop:duration-order}
For integers $1\le d\le D$, the embedding $f(s)=1+\lfloor(s-1)D/d\rfloor$ preserves both benchmarks and gives $\rho_{\mathcal B}(d)\ge \rho_{\mathcal B}(D)$.
\end{proposition}
\emph{Proof.}
Place request $s$ at $f(s)$ on the $D$-clock and give every intervening slot an empty menu. The map $f$ is strictly increasing and satisfies $f(s+d)=f(s)+D$. If $u-s\ge d$, then $f(u)-f(s)\ge D$. If $0<u-s<d$, monotonicity gives $f(u)<f(s+d)=f(s)+D$. Thus two original assignments to the same resource conflict exactly when their images do. This proves equality of feasible integer schedules and preservation of their rewards.

At an embedded arrival $f(t)$ the backward $D$-window contains exactly the images of the original backward $d$-window. The same fractional variables therefore satisfy the two sets of occupancy constraints. An intervening empty slot adds no variable, and its load is at most the load at the most recent embedded arrival. Hence the complete window LP also has the same objective values.

For each known $(I,w)$ simulate a $D$-policy by supplying the intervening empty menus before revealing the next real menu. Its past revealed inputs and private randomness determine every simulated decision. The resulting $d$-policy is online, has the same reward pathwise and inherits the $D$-policy's common-prefix law. Every guarantee attainable at $D$ is consequently attainable at $d$. Taking suprema proves the claim.\hfill$\square$

The fixed-duration values differ from the uniform guarantee in Theorem~\ref{thm:main}. At $d=1$ a maximum-weight choice at each request is optimal, so $\rho_{\mathcal B}(1)=1$. Theorem~\ref{thm:fixed-two} gives $\rho_{\mathcal B}(2)=3/4$. For general $d$, Theorem~\ref{thm:untrimmed} and the $n=d$ permutation instance in Section~\ref{sec:upper} give
\[
 1-1/e+\frac{2^{-21}}{d^2}\ \le\ \rho_{\mathrm{LP}}(d)\ \le\ \rho_{\mathrm{OPT}}(d)
 \ \le\ \min\{1,(1-1/e)(1+1/d)\}.
\]
Both benchmarks equal $d$ on that permutation instance. Feige's analysis of integral online algorithms on the permutation distribution \cite[Theorem~6 in the full version]{feige} sharpens its upper bound to
\[
 \rho_{\mathrm{OPT}}(d)\le 1-1/e+\frac{1-2/e}{d}+O\!\left(\frac{1}{d\,d!}\right).
\]
Thus both optimal ratios are nonincreasing and converge to $1-1/e$. Their exact values, and whether they coincide, remain open for general fixed $d\ge3$.

\subsection{Exact ratio for duration two}\label{app:fixed-two}

At $d=2$ the window LP is integral by Madarasi \cite[Theorem~4]{madarasi}: his ordered left vertices are the slots, and each resource has unit capacity on consecutive pairs.

\begin{theorem}[Exact optimal ratio at duration two]\label{thm:fixed-two}
For a finite resource set $I$ with known nonnegative weights $w$ and at least one positive weight, suppose that each complete menu, including busy resources, is revealed on arrival. Among randomized online policies that know $(I,w)$ and use the same rule on every horizon, the optimal competitive ratio at $d=2$ against either offline benchmark is $r=1-q$, where
\[
 q=\max_{\substack{i\ne j\\ w_i,w_j>0}}\frac{w_iw_j}{(w_i+w_j)^2}
\]
and \(q=0\) when there is only one positive entry. A single online policy for \(w\) attains \(r\) on every finite horizon. The worst ratio over weight vectors is \(3/4\).
\end{theorem}

For example, let the weights be $(1,2)$ and offer both resources followed by only the heavier one. Under delayed restoration, given the light resource's rank $u$, it wins the first request with probability $-\log((1+e^{-u})/2)$. The heavier resource earns two units of reward in either case. The lighter one earns an additional unit exactly when it wins the first request. Dividing the expected reward $2+\int_0^1-\log((1+e^{-u})/2)\,du$ by the offline optimum of three gives a ratio of approximately $0.73644$. The weight-specific policy in the theorem guarantees $7/9$ on every input with these weights. Thus delayed restoration attains the uniform constant and its strict fixed-duration improvement, while another policy can achieve a larger guarantee at a fixed duration.

The policy tracks the optimal prefix reward $F$ and the loss $\lambda_i$ from forbidding $i$ as the final action. At the next menu the two largest effective scores $w_i-\lambda_i$ determine the new optimum and its vector of losses. The policy randomizes between those two resources, taking account of which one it occupied in the previous slot. If $p_i$ is the resulting probability of occupying $i$ and $G$ the expected prefix reward, it maintains
\[
 G-rF\ge\max\big\{0,\ \max_i(w_i p_i-qw_i-r\lambda_i)\big\}.
\]
The first bound guarantees the claimed ratio at every stopping horizon fixed by the input. The second reserves enough reward for a possible singleton request for $i$ in the next slot. Appendix~\ref{app:two-proof} gives the policy and verifies that its required marginals can always be realized.

For the upper bound, take two distinct resources with positive weights \(a_1,a_2\) and offer both in the first slot. Offer only the resource of weight \(a_1\) in the second slot with probability \(a_1/(a_1+a_2)\), and only the other resource otherwise. Both inputs have offline value \(a_1+a_2\). Either first match gives expected total reward at most \(a_1+a_2-a_1a_2/(a_1+a_2)\), and idling gives at most \((a_1^2+a_2^2)/(a_1+a_2)\). The two sequences are fixed independently of the policy's random choices, so averaging over the input and the policy's randomness yields a fixed sequence with ratio at most \(1-a_1a_2/(a_1+a_2)^2\). Take a pair attaining \(q\). With one positive weight the offline benchmark already bounds the ratio by one. The worst value is \(3/4\) because \(a_1a_2/(a_1+a_2)^2\le1/4\), with equality at \(a_1=a_2\). The equal-weight two-request instance also appears in the lower bound for ordinary degree-two fractional matching \cite[Theorem~12]{bch}.
\hfill\(\square\)

\section{Conclusion}\label{sec:conclusion}
Delayed restoration attains $c=1-1/e$ against the full window LP for unit inventories, known nonnegative resource weights and a common deterministic service duration. An earlier use of a resource supplies the credit needed when that resource is busy at a later request. Ordinary matching shows that this guarantee is sharp uniformly over durations. With arbitrary arrival density, the same upper bound applies at each prescribed positive real duration.

For the integer clock, fixing the duration changes the problem. The optimal ratios against the integer optimum and the window LP are nonincreasing in $d$. Both equal $3/4$ at $d=2$, while for general $d$ the proved lower gain is $2^{-21}/d^2$ and the classical permutation distribution gives an $O(1/d)$ upper gain. These bounds motivate the following question about the best policy.
\begin{conjecture}[First-order gain]\label{conj:first-order}
With unit inventories, known nonnegative resource weights, complete revealed menus and at most one indivisible request per integer slot,
\[
 \rho_{\mathrm{LP}}(d)=1-1/e+\Theta(1/d)
 \qquad\text{as }d\to\infty.
\]
\end{conjecture}
The permutation upper bound gives the $O(1/d)$ direction. For the matching lower bound one needs a universal $a>0$ and $d_0$ such that, for every $d\ge d_0$ and every known resource set and weight vector, some online policy guarantees $(1-1/e+a/d)\mathrm{LP}$ on every input. The conjecture allows the full current menu, as in the duration-two policy. Whether the same order is attainable from free offers alone, and the order attained by delayed restoration itself, remain open.

\section*{Supplementary material}
Appendix~\ref{app:formal-statements} states the Lean theorems for the integer scheduling problem and the window LP and identifies the results covered by kernel checks. We used AI tools to assist with the Lean verification. The proofs in the article establish the guarantees directly. The accompanying sources and verification records permit a separate check of the formal statements.

\clearpage
\appendix
\renewcommand{\theHsection}{appendix.\Alph{section}}
\section{State spaces and conditional kernels}\label{app:window-measurability}
This appendix gives the measurable constructions used in Section~\ref{sec:product}. Fix the finite resource set and a finite input. For each resource, represent its active region and each pending region by sorted, disjoint left-open, right-closed intervals. Store their endpoints in $[0,1]$, their positive integer deadlines, their physical release times and, when free, a retained rank in $[0,1]$. For any prescribed number of intervals and discrete calendar entries, the admissible endpoint vectors form a Borel subset of a finite-dimensional Euclidean space: all order, disjointness and deadline conditions are finite systems of strict or weak inequalities. The full space is a countable disjoint union of these Borel spaces, and is therefore standard Borel. A fixed public label order makes every winning choice a measurable function. Union, intersection and clipping are piecewise continuous in sorted endpoints, with Borel pieces selected by comparisons. The calendar and reset transitions are consequently Borel.

\paragraph{Joint measurability of the inverse sampler.}
If $A=\bigcup_{a=1}^k(l_a,u_a]$ is represented by disjoint intervals and $z>0$, then
\[
 -\log F_A(z)
 =\sum_{a=1}^k
  {\bf1}\{u_a>\max(l_a,z)\}
  \log\frac{u_a}{\max(l_a,z)}.
\]
Every summand is Borel in its endpoints and $z$. Define $F_A(0)=\lim_{n\to\infty}F_A(1/n)$, which is Borel as a pointwise limit. The generalized inverse
\[
 Q(A,u)=\inf\{z\in[0,1]:F_A(z)\ge u\},\qquad 0<u<1,
\]
is jointly Borel in $(A,u)$. Indeed, for $a\in(0,1]$,
\[
 \{(A,u):Q(A,u)<a\}
 =\bigcup_{q\in\mathbb Q\cap[0,a)}\{(A,u):u\le F_A(q)\}.
\]
Right continuity of a cumulative distribution justifies this identity. Taking a fresh uniform coordinate gives the law $M(A)$, including its possible atom at zero. A countable product of independent unit uniforms realizes every finite- or infinite-horizon sampler draw and every regional restoration draw.

\paragraph{Conditional laws at observed prices.}
An observation at a slot belongs to the disjoint union of a rejection point and $I_+\times[0,\max_i V_i(1)]$. Finite observation histories are standard Borel. The joint state and observation laws therefore admit regular conditional probabilities. In Lemma~\ref{lem:conditional}, the displayed posterior $Q_{j,b}$ is Borel in the predecision state and $b$: its numerator is the integral of the bounded Borel function ${\bf1}_{C}(z){\bf1}_{V_j(z)\le b}$ against the Borel state-dependent kernel $M(A_j)$, and its denominator is the same integral without ${\bf1}_C$. On the zero-denominator Borel set assign the point mass at zero. The product kernel verifies the disintegration identity by cancellation in \eqref{eq:winner-factor}. The monotone class theorem extends the identity from rectangles to all Borel events. Conditional statements in the proof are made on the resulting probability-one set of boundary histories. For a fixed finite input there are finitely many boundaries and target edges, so their exceptional sets can be combined into a single null set when needed.

\paragraph{Measurable Poisson paths.}
A unit-rate Poisson process on $[0,\infty)$ has a standard Borel realization by its increasing sequence of arrival times. Transforming its points by $x\mapsto e^{-x}$ gives intensity $dz/z$ on $(0,1]$. Each compact subinterval bounded away from zero contains finitely many points almost surely. The maximum in a finite union $B$ of intervals is measurable, jointly in the interval endpoints and the point sequence, because
\[
 \{\max(\Pi\cap B)\le z\}
 =\bigcap_{n\ge1}\{\pi_n\notin B\cap(z,1]\}.
\]
Independent restrictions to disjoint regions follow from independent Poisson increments. For every region containing an interval next to zero the maximum is positive almost surely. In a finite input all regions used by the construction are finite unions of intervals; all maxima, finite comparisons and state updates in the maps $\Phi_S$ are measurable. Their pushforward laws therefore provide the conditional kernels asserted in Lemma~\ref{lem:terminal}.

\paragraph{Observable state and full sampler history.}
The persistent representation conditions on the recorded winners and prices, which retain the rank truncations described in Lemma~\ref{lem:conditional}. If a history instead reveals all previously sampled ranks, a surviving old rank is generally determined and has a different posterior. One can still restart the fresh sampler from its observable state: future uniform coordinates are independent of its full past, and Corollary~\ref{lem:fresh-law} identifies the future observation kernel from that state. The window proof uses the first conditioning throughout.

\section{Embedding a common release calendar}\label{sec:calendar}
Consider a common first-in, first-out (FIFO) release calendar. At arrival $t$, a timer started at $s<t$ has expired exactly when $s\le k_t$, for a nondecreasing sequence $0\le k_t<t$. The calendar and menus are fixed independently of the policy's coins. The scheduler observes the identities of its completed matches before each decision, together with the free eligible resources. It need not know $k_t$, future deadlines, or the eligibility of busy resources.

\begin{proposition}\label{prop:calendar}
Given the free offers and the identities of all its own completed matches, including returns absent from the current menu, delayed restoration earns in expectation at least $1-1/e$ times the fractional occupancy optimum. The constant is sharp over this class of calendars.
\end{proposition}
\emph{Proof.} First fix the entire calendar. Compare with an executor given its announcements, using identical private draws. At a match, attach every new exclusion to the winning job. Release all surviving exclusions in its bundle when that job completes. An intervening use of a loser discards its old exclusions, exactly as in Algorithm~\ref{alg:continuous}; tag entries by assignment generation to ignore such stale records. A rejection creates no exclusions. Each live entry and its winning job have the same origin and therefore expire together. Induction gives identical active regions, available offers, winners and prices in the two executions. This reasoning uses completion notifications even when a returning resource is absent from the current menu. 
It remains to transfer the integer-clock guarantee. Delong et al.\ \cite[Appendix A]{delong} embed constant real durations by inserting empty arrivals. Here the cutoff sequence itself determines such an embedding. For $T\ge1$ arrivals we construct increasing integers $f(t)\le T^2$ with
\[
    s\le k_t\quad\Longleftrightarrow\quad f(t)-f(s)\ge T \qquad(s<t).
\]
First choose real positions $z_1=0<\cdots<z_T$ satisfying the same equivalence with threshold one. If $k_t=t-1$, set $z_t=z_{t-1}+1$. Otherwise choose $z_t$ halfway between $\max\{z_{t-1},z_{k_t}+1\}$ and $z_{k_t+1}+1$, omitting $z_{k_t}+1$ when $k_t=0$. Put $u=k_t+1$. If $u=t-1$, then $z_u+1>z_{t-1}$. If $u<t-1$, monotonicity gives $u>k_{t-1}$, so the induction hypothesis gives $z_{t-1}-z_u<1$. Thus the upper endpoint $z_u+1$ exceeds $z_{t-1}$ in both cases. The endpoint also exceeds $z_{k_t}+1$ when present. Thus the new position preserves exactly the specified conflicts, and every increment is at most one. Write $z_t=m_t+a_t$ with integer part $m_t$ and fractional part $a_t$. Rank the distinct fractional parts from zero in increasing order and put $f(t)=1+Tm_t+\operatorname{rank}(a_t)$. When $m_t-m_s=1$, both threshold tests reduce to $a_t\ge a_s$. Integer-part differences zero or at least two settle the other cases. Since $z_T\le T-1$, the stated size bound follows.

Place each menu at its image and give every intervening slot an empty menu. At a real arrival exactly the same origins have expired. Empty slots create no use or exclusion. Coupling the draws at corresponding arrivals therefore preserves every winner and price. The occupancy constraints at the images agree, and between them the fractional load can only decrease. Hence the two LPs have equal value, and Theorem~\ref{thm:main} at duration $T$ gives the claim. The embedding is used only to compare fixed inputs; the policy operates from the observed completions. Calendars with $k_t=0$ throughout include ordinary matching and give sharpness.\hfill$\square$

The order condition matters even for a single unit-weight resource with advance notice of each job's duration. A first job spans $m+1$ slots. Either it is the only offer or it is followed by $m$ unit-slot jobs. A policy accepting it with probability $p$ has ratios at most $p$ and $1-p+p/m$, respectively. Maximizing their minimum gives $m/(2m-1)$, which tends to $1/2$. For a fixed horizon the shorter input can be padded with empty menus.

\section{The duration-two policy}\label{app:two-proof}
For Theorem~\ref{thm:fixed-two}, let $F$ and $G$ denote the offline and expected online prefix rewards. Let $\lambda_i$ be the offline loss from forbidding $i$ as the final action, and $p_i$ its online probability of being the last action. These quantities are determined by the revealed menus. The policy also records its realized last action $a_{\rm prev}$.
Put $r=1-q$ as in Theorem~\ref{thm:fixed-two} and write $\xi=G-rF$. The policy maintains the following prefix invariant
\begin{equation}\label{eq:fixed-two-invariant}
 \xi\ge0,\qquad \xi\ge w_i p_i-qw_i-r\lambda_i \quad\text{for every }i.
\end{equation}

\begin{algorithm}[ht]
\caption{One-slot marginal policy for $d=2$}\label{alg:one-slot}
\begin{algorithmic}[1]
\State Initialize $F=G=0$, $\lambda=p=0$ and the realized last action $a_{\rm prev}$ to idle.
\For{each arriving menu $S$ of positive-weight resources}
  \State Set $\xi=G-rF$. Rank $i\in S$ by $w_i-\lambda_i$ with the fixed tie order.
  \State Set $p'=\lambda'=0$, $u=0$ and the current action $a_{\rm new}$ to idle.
  \If{$S=\{i\}$}
    \State Set $u=\lambda'_i=w_i-\lambda_i$, $p'_i=1-p_i$. Choose $a_{\rm new}=i$ if $a_{\rm prev}\ne i$.
  \ElsIf{$|S|\ge2$}
    \State Take the top two resources $i,j$ with scores $u,v$, and set $\lambda'_i=u-v$.
    \State Set $p'_i=x=\max\{p_j,0,(ru-qw_j-\xi)/w_i\}$ and $p'_j=1-x$.
    \State If $a_{\rm prev}=i$, set $a_{\rm new}=j$. If $a_{\rm prev}=j$, set $a_{\rm new}=i$.
    \State For $a_{\rm prev}\notin\{i,j\}$ set $a_{\rm new}=i$ with probability $(x-p_j)/(1-p_i-p_j)$ and $a_{\rm new}=j$ otherwise (choose $j$ if the denominator vanishes).
  \EndIf
  \State Set $(F,G,\lambda,p,a_{\rm prev})\gets(F+u,G+\sum_k w_k p'_k,\lambda',p',a_{\rm new})$.
\EndFor
\end{algorithmic}
\end{algorithm}

\emph{Proof.}
Deleting a final use of $i$ gives $0\le\lambda_i\le w_i$. Choosing $i$ now yields offline value $F+w_i-\lambda_i$, while idling gives $F$. Thus $F$ increases by the highest score $u$, and forbidding its resource costs the gap $u-v$ to the runner-up. All other losses vanish. Use zero for a missing score. This proves the offline updates in Algorithm~\ref{alg:one-slot}.

The inequalities hold initially when \(G=0\) and \(p_i=0\). An empty request leaves \(\xi\) unchanged and resets all \(p_i,\lambda_i\) to zero. On a singleton request the policy uses \(i\) whenever available. Its new marginal is \(p'_i=1-p_i\). With \(u=w_i-\lambda_i\), the old invariant gives \(w_i(1-p_i)\ge ru-\xi\), so \(\xi'=\xi+w_i p'_i-ru\ge0\). The new inequality for \(i\) reduces to \(\xi\ge-qw_i\), using \(\lambda'_i=u\). The other inequalities follow from \(\xi'\ge0\).

For at least two eligible resources, let \(i,j,u,v\) be as above and put \(a_1=w_i\), \(a_2=w_j\). Define
\[
 X=\max\{0,(ru-qa_2-\xi)/a_1\},\qquad
 Y=\max\{0,(rv-qa_1-\xi)/a_2\}.
\]
Choose new marginals \(x=\max\{p_j,X\}\) for \(i\) and \(y=1-x\) for \(j\). To verify feasibility, the old invariant gives \(a_1(1-p_i)\ge ru-\xi\) and \(a_2(1-p_j)\ge rv-\xi\). Hence \(X\le1-p_i\) and \(Y\le1-p_j\). Also \(p_i+p_j\le1\) and each of \(X,Y\) is at most \(r\). If both are positive, substitute $u\le a_1$, $v\le a_2$ and $\xi\ge0$ in their definitions. Since $q\ge a_1a_2/(a_1+a_2)^2$, this gives
\[
 X+Y\le 2r-q(a_2/a_1+a_1/a_2)=2-q\frac{(a_1+a_2)^2}{a_1a_2}\le1.
\]
If either is zero the sum is at most \(r\). These bounds give \(p_j\le x\le1-p_i\), \(x\ge X\), and \(y\ge Y\).

The realized action must switch after $i$ or $j$. The remaining histories have mass $1-p_i-p_j$ and must supply mass $x-p_j$ to $i$. Since $p_j\le x\le1-p_i$, the conditional lottery in Algorithm~\ref{alg:one-slot} realizes both marginals, including when the remaining mass is zero.

The expected reward increment is \(a_1x+a_2y\). If \(a_1\le a_2\) this is at least \(a_1\ge u\). If \(a_1>a_2\), the reward is increasing in \(x\). At \(x=X\) it is at least \(ru-\xi\) because \(a_1X\ge ru-qa_2-\xi\) and \(1-X\ge q\). Thus \(\xi'=\xi+a_1x+a_2y-ru\ge0\). The new inequalities in \eqref{eq:fixed-two-invariant} for \(i\) and \(j\) reduce to \(a_2y\ge rv-qa_1-\xi\) and \(a_1x\ge ru-qa_2-\xi\), respectively. They hold by \(y\ge Y\) and \(x\ge X\). For each remaining resource, \(p'_k=\lambda'_k=0\). Induction gives \(G\ge rF\) at every horizon.

This proves the lower bound for either benchmark by integrality of the duration-two window LP.\hfill$\square$

\section{Formal statements and verification scope}\label{app:formal-statements}
The declarations below are stated mathematically with their Lean identifiers. Slots in the sources are indexed from zero. A finite roster $C\subseteq\mathbb N$ represents the resource set, and a schedule is a map $p:\{0,\ldots,T-1\}\to C\cup\{\varnothing\}$ (\texttt{Schedule}). Its reward is $R_w(p)=\sum_{t:p(t)\ne\varnothing}w_{p(t)}$ (\texttt{scheduleReward}). The predicate \texttt{LegalSchedule} requires $p(t)\in N_t$ whenever $p(t)\ne\varnothing$, and $t-s\ge d$ whenever $s<t$ and $p(s)=p(t)\ne\varnothing$. The function \texttt{offlineOPT} is the maximum of $R_w$ over this finite, nonempty set of legal schedules; the idle schedule establishes nonemptiness.

\paragraph{The original integer-clock question.}
The declaration \nolinkurl{MOR93.CompleteOR93.complete_OR93_integer} proves both of the following statements. For every finite $C$, every $w$ nonnegative on $C$, and every integer $d>0$, there are measures $L_{N,T}$ on schedules, simultaneously for all menu sequences $N$ and all finite $T$, such that
\begin{equation}\label{eq:formal-original}
 \begin{gathered}
 L_{N,T}\text{ is a probability measure},\qquad
 L_{N,T}(\{p:p\text{ is legal}\})=1,\\
 \int R_w(p)\,dL_{N,T}(p)\ge(1-e^{-1})\mathrm{OPT}(C,w,N,d,T),\\
 N_s=N'_s\ (s<k),\quad k\le T,U
 \ \Longrightarrow\
 (\pi_{k,T})_\#L_{N,T}=(\pi_{k,U})_\#L_{N',U}.
 \end{gathered}
\end{equation}
Here \texttt{takeSchedule}, denoted by $\pi_{k,T}$, restricts a schedule to its first $k$ slots. The symbol $\#$ denotes pushforward. The last equality requires the same prefix law across horizons and all future continuations. Since the action set is finite, these laws define a causal randomized online policy.

For the upper bound, \texttt{OriginalFamily}$\,(n)$ consists of probability measures on schedules for $C_n=\{0,\ldots,n\}$, at duration $n+1$, with almost-sure legality and exactly the prefix property above. The declaration quantifies over every such family, without a competitive lower-bound hypothesis. For every $\varepsilon>0$ it gives an $n$ such that for every \texttt{OriginalFamily}$\,(n)$ there exists a permutation $\sigma$ of $C_n$ with
\[
 \int R_{\boldsymbol1}(p)\,dL_{N^\sigma,n+1}(p)
 \le(1-e^{-1}+\varepsilon)\mathrm{OPT}(C_n,\boldsymbol1,N^\sigma,n+1,n+1).
\]
The menu $N^\sigma_s=C_n\setminus\{\sigma(0),\ldots,\sigma(s-1)\}$ is \texttt{suffixMenu}. Thus the lower statement answers the original existence question independently of a correspondence between a particular implementation and Algorithm~\ref{alg:continuous}.

\paragraph{The specified policy and the window LP.}
The predicate \texttt{FeasibleLP} holds for arrays $x_{it}$, $i\in C$, $0\le t<T$, that are nonnegative, vanish off $i\in N_t$, and satisfy
\[
 \sum_{i\in C}x_{it}\le1,\qquad
 \sum_{\substack{0\le s\le t\\t<s+d}}x_{is}\le1.
\]
These inequalities hold for each request and each resource in the roster, respectively. Values at $i\notin C$ are immaterial. The definition \texttt{windowOptimum} is the supremum of $\sum_{t,i\in C}w_ix_{it}$ over these arrays. For all $C,w,N,d,T$ with $w_i\ge0$ on $C$ and $d>0$,
\nolinkurl{MOR93.TracePacking.review_integer_policy} states that \texttt{actualLaw} is a probability measure on histories $h_t=(q_t,b_t)$, almost surely assigns only resources in the roster and the revealed menu and separates successive uses by at least $d$ slots, and satisfies
\[
 (1-e^{-1})\mathrm{LP}(C,w,N,d,T)
 \le\int\sum_{t=0}^{T-1}\texttt{rewardAt}(w,h_t)\,d\texttt{actualLaw}(h).
\]
The code $q_t=0$ denotes rejection and $q_t=i+1$ denotes resource $i$; \texttt{rewardAt} is respectively zero or $w_i$. The law starts with all resources free, draws independent uniforms through \texttt{drawRegion}, chooses the maximal bid through \texttt{liveWinner}, and updates readiness and return times through \texttt{advance}. Its active set is $\{z:0<z\le1,\ \texttt{readyAt}(i,z)\le t\}$. The declarations \texttt{calendar\_refines\_timers}, \texttt{active\_regions\_equal} and \texttt{same\_winner} in \nolinkurl{MOR93.IntervalCalendar} equate these updates, active sets, and winner--price pairs with finite interval calendars under the same uniforms.

\paragraph{Finite ranks and coverage.}
For each integer $m>0$, \nolinkurl{OR93Independent.FiniteFlags.Clocks.finite_policy_LP} proves the full window-LP guarantee
$1-(m/(m+1))^m$ for the finite scan. At slot $s$, scan the active pairs $(i,k)$ of free offered resources, $1\le k\le m$, in decreasing order of bid $w_i[1-(m/(m+1))^k]$, breaking ties by resource label and then level, both in decreasing order. Each test succeeds independently with probability $1/k$; stop at the first success and match its resource, or reject if the scan is exhausted. Set the return time of the winning resource, all its levels, and each tested pair that failed to $s+d$. The declaration \nolinkurl{MOR93.TapeClosure.actual_grid_fractional} transfers this bound to the uniform-tape executor. These claims cover arbitrary fixed finite menus, nonnegative resource weights and common integer $d>0$.

The sources also prove the clock transfer, integer-duration monotonicity and duration-two bounds against integer $\mathrm{OPT}$. The duration-two result was first checked against imported dependencies and subsequently rebuilt from its 19 source modules with a fresh kernel check of the complete import closure. The core, extension, finite-grid, common-tape, monotonicity and interval-calendar records also include fresh kernel checks. The local aggregation record covers algebra and finite counting. The probability estimates of Theorem~\ref{thm:untrimmed} are proved analytically in Section~\ref{sec:gain}; the LP-integrality step at $d=2$ uses Madarasi~\cite{madarasi}. All reported theorem-axiom lists contain only \texttt{propext}, \nolinkurl{Classical.choice} and \nolinkurl{Quot.sound}.

\end{document}